\documentclass[11pt]{article}
\pdfoutput=1
\usepackage[centertags]{amsmath}
\usepackage[square, comma, sort&compress,numbers]{natbib}
\usepackage{array,multirow}
\numberwithin{equation}{section}
\usepackage{amssymb,amsfonts}
\usepackage{graphicx}
\usepackage{color}
\usepackage{mathtools,bm}
\usepackage{accents}

\usepackage{epsfig}
\usepackage{bbold}

\usepackage{wrapfig}
\usepackage{float}
\usepackage{soul}
\usepackage{tkz-euclide}

\usepackage{tikz,pgf}
\usetikzlibrary{shapes}
\usetikzlibrary{calc}
\usetikzlibrary{decorations.pathmorphing}
\usetikzlibrary{decorations.pathreplacing,shapes.misc}
\usetikzlibrary{positioning}
\usetikzlibrary{arrows}
\usetikzlibrary{decorations.markings}
\usetikzlibrary{shadings}

\usetikzlibrary{intersections}

\def\be{\begin{equation}}
\def\ee{\end{equation}}
\def\ba{\begin{array}}
\def\ea{\end{array}}

\def\dps{\displaystyle}

\def\tr{{\rm Tr}}

\def\1{\tilde{1}}
\def\2{\tilde{2}}
\def\3{\tilde{3}}

\newdimen\tableauside\tableauside=1.0ex
\newdimen\tableaurule\tableaurule=0.4pt
\newdimen\tableaustep
\def\phantomhrule#1{\hbox{\vbox to0pt{\hrule height\tableaurule
width#1\vss}}}
\def\phantomvrule#1{\vbox{\hbox to0pt{\vrule width\tableaurule
height#1\hss}}}
\def\sqr{\vbox{%
\phantomhrule\tableaustep

\hbox{\phantomvrule\tableaustep\kern\tableaustep\phantomvrule\tableaustep}%
\hbox{\vbox{\phantomhrule\tableauside}\kern-\tableaurule}}}
\def\squares#1{\hbox{\count0=#1\noindent\loop\sqr
\advance\count0 by-1 \ifnum\count0>0\repeat}}
\def\tableau#1{\vcenter{\offinterlineskip
\tableaustep=\tableauside\advance\tableaustep by-\tableaurule
\kern\normallineskip\hbox
{\kern\normallineskip\vbox
{\gettableau#1 0 }%
\kern\normallineskip\kern\tableaurule}%
\kern\normallineskip\kern\tableaurule}}
\def\gettableau#1 {\ifnum#1=0\let\next=\null\else
\squares{#1}\let\next=\gettableau\fi\next}

\newcommand{\bref}[1]{\textbf{\ref{#1}}}

\newcommand{\im}{\mathop{\mathrm{Im}}}

\def\cC{\mathcal{C}}
\def\cD{\mathcal{D}}

\def\cF{\mathcal{F}}

\def\cH{\mathcal{H}}

\def\cL{\mathcal{L}}

\def\cN{\mathcal{N}}
\def\cO{\mathcal{O}}

\def\cV{\mathcal{V}}

\numberwithin{equation}{section} \makeatletter
\@addtoreset{equation}{section}

\def\ads{AdS$_{3}\;$}
\def\be{\begin{equation}}
\def\ee{\end{equation}}
\def\ba{\begin{array}}
\def\ea{\end{array}}

\def\dps{\displaystyle}

\def\ba{\begin{array}}
\def\ea{\end{array}}

\def\dps{\displaystyle}

\def \bz{\bar z}

\def\cft{CFT$_2$ }
\def\ads{AdS$_3$ }
\def\adscft{AdS$_3$/CFT$_2$ }

\def\bx{{\bf x}}

\def\bz{{\bf z}}

\newcommand{\dl}{h}
\newcommand{\td}{\tilde{h}}

\newcommand{\bh}{\bar{h}}

\newcommand{\cas}{\mathbb{C}_2[n,j]}

\def\Tr{\operatorname{Tr}}
\def\C2{\text{C}_2}
\def\torus{\mathbb{T}^2}

\usepackage{jheppub}
\makeatletter
\def\@fpheader{\vspace{-.1cm}}
\makeatother

\title{\centering{Torus conformal blocks and Casimir equations\\ in the necklace channel   }}

\author{Konstantin\ Alkalaev,}
\author{Semyon\ Mandrygin,}
\author{Mikhail\ Pavlov}

\affiliation{I.E. Tamm Department of Theoretical Physics, \\P.N. Lebedev Physical
Institute, 119991 Moscow, Russia}

\emailAdd{alkalaev@lpi.ru}
\emailAdd{semyon.mandrygin@gmail.com}
\emailAdd{pavlov@lpi.ru}

\abstract{We consider the conformal block decomposition in arbitrary  exchange channels of a two-dimensional  conformal field theory on a torus. The  channels are described by diagrams built of a closed loop  with external legs (a necklace sub-diagram) and trivalent vertices forming trivalent trees attached to the necklace.  Then, the $n$-point torus conformal block in any channel can be  obtained by acting with a  number of OPE operators  on the $k$-point  torus block in the necklace channel at  $k=1,...,n$. Focusing on the necklace channel, we go to the large-$c$ regime, where the Virasoro algebra truncates to the $sl(2, \mathbb{R})$ subalgebra, and obtain  the system of the Casimir equations for the respective  $k$-point global conformal block.  In the plane limit, when the torus modular parameter $q\to 0$, we explicitly find the Casimir equations on a plane which define  the $(k+2)$-point global conformal block in the comb channel. Finally, we formulate the general scheme to find Casimir equations for global torus blocks in arbitrary channels.}

\begin{document}

\maketitle
\flushbottom

\section{Introduction}

Correlation functions of conformal field theory (CFT) associated with a particular algebra of local/global conformal symmetries can be decomposed into conformal blocks which are completely fixed by the conformal symmetry. The conformal blocks are instrumental in the conformal/modular  bootstrap programs \cite{Belavin:1984vu,Verlinde:1988sn} (for recent developments see e.g.  \cite{Poland:2018epd,Collier:2016cls,Hartman:2022zik,Bissi:2022mrs}). The deep role of conformal blocks was revealed in the context of AdS/CFT correspondence, where they are  dual to lengths of geodesic diagrams stretched in the bulk \cite{Hartman:2013mia,Fitzpatrick:2014vua,Hijano:2015rla,Alkalaev:2015wia,Hijano:2015zsa,Hijano:2015qja,Banerjee:2016qca}.

Two-dimensional conformal field theories (CFT$_2$) on Riemannian surfaces are interesting from different perspectives. The original motivation comes from the study of  critical phenomena in statistical physics and string theory amplitudes, see e.g. \cite{Cardy:1986ie,Itzykson:1986pj,Eguchi:1986sb}. The first cases of Riemannian surfaces --  sphere $\mathbb{S}^2$ and torus $\mathbb{T}^2$  -- are both ubiquitous and rather simple. On the other hand, while the sphere (plane) \cft is  quite well--studied, this is not so for the torus CFT$_2$, especially, when it comes to multipoint correlation functions and corresponding conformal blocks.\footnote{For a general formulation of \cft on Riemannian surfaces see e.g. \cite{Itzykson:1986pj,Eguchi:1986sb}. Literature on multipoint ($n\geq2$) torus blocks is rather scarce: $n$-point  torus blocks in the necklace channel were studied within the recursive representation \cite{Cho:2017oxl}; various large-$c$ torus blocks  were studied from \adscft perspective in   \cite{Kraus:2017ezw,Alkalaev:2017bzx,RamosCabezas:2020mew,Gerbershagen:2021yma}.} The large-$c$ regime can significantly simplify the form of various conformal functions which are often not even known for arbitrary values of the conformal dimensions and the central charge. On the other hand, by the Brown-Henneaux formula $c \sim 1/G_N$, a large-$c$ \cft  is holographically dual to \ads quantum gravity  in the semiclassical approximation \cite{Brown:1986nw}.

In the torus \cft already $1$-point conformal block is not known in a closed form. The {\it global} 1-point torus block (which is associated to the $sl(2, \mathbb{R})\subset Vir$ subalgebra) is found to be a hypergeometric function \cite{Hadasz:2009db}. As in any CFT with a finite-dimensional symmetry group, the global blocks can be equivalently described as eigenfunctions of the Casimir operators associated with irreducible conformal families in the exchange channels \cite{Dolan:2003hv,Dolan:2011dv} (see e.g. \cite{Bissi:2022mrs} for recent review). In the case of the torus \cft the study of the Casimir approach was initiated in  \cite{Kraus:2017ezw}, where the  Casimir equation for the 1-point torus block was found to be a 2nd order differential (hypergeometric) equation in the modular parameter.\footnote{This result naturally extends to $n$-point blocks in  the so-called OPE channel \cite{Kraus:2017ezw}.  The same technique was used to compute thermal 1-point conformal blocks in CFT$_d$ \cite{Gobeil:2018fzy} and torus  1-point $\cN=1$  superblocks \cite{Alkalaev:2018qaz}.}

In this paper, we analyze multipoint torus blocks in general channels with a focus on the so-called  necklace channel. This channel is represented by  a closed loop with a number of external legs and  plays the role of a building block of any other torus diagram. Going to the large-$c$ limit, we formulate the Casimir equations for {\it global} torus blocks in the necklace channel.

Note that from the plane \cft perspective,  the $n$-point torus block  can be understood by gluing  the endpoints of  the plane $(n+2)$-point block. In the plane limit, when the torus modular parameter  $q \rightarrow 0$, the leading asymptotics of the torus block in the necklace channel defines the plane block in the comb channel (see, e.g. \cite{Cho:2017oxl}).\footnote{Multipoint blocks in the comb channel were explicitly computed in \cite{Rosenhaus:2018zqn} (see also \cite{Fortin:2019zkm}).} We show that in this case the limiting Casimir torus equations coincide with the Casimir plane equations and, therefore, find explicit realization of the Casimir operators of the plane CFT$_2$.

Since  both sphere and torus global blocks can be considered as $c=\infty$  limiting functions they contain much less information than the full Virasoro conformal blocks. Nonetheless, they still have a wide range of applications related to $1/c$ calculations, especially, in the context of \adscft correspondence. E.g., the global blocks are directly  related to the classical blocks \cite{Zamolodchikov1986} with light ($\Delta \sim \cO(c^0)$) and heavy ($\Delta \sim \cO(c^1)$) operators through a chain of relations \cite{Fitzpatrick:2015zha,Alkalaev:2015fbw,Alkalaev:2016fok,Alkalaev:2018qaz}. The important role of the heavy-light classical conformal blocks is that they calculate the lengths of geodesic networks stretched in asymptotically \ads spaces. Also, there is  a convenient representation of  Virasoro conformal blocks as a sum over global blocks which finds fruitful application in $1/c$ treatment of many problems including the entanglement entropy calculation and the scattering problem in \ads  \cite{Perlmutter:2015iya}. The same useful representation should have been found in torus CFT$_2$. In this case, however, the  global torus blocks are to be substituted by the so-called  light torus blocks \cite{Alkalaev:2016fok}, which are in their turn are related to torus blocks by a simple transformation.

The paper is organized as follows. In the beginning of Section \bref{sec:def} we fix our notation and conventions related to the torus CFT$_2$. In  Section \bref{sec:correlation} we review multipoint torus correlation functions. In Section \bref{sec:block_diag} we consider   the general structure of   the conformal block decomposition and introduce a classification of torus diagrams by a number of external legs of the necklace sub-diagram. Recognizing the necklace sub-diagram as the basic constituent, any multipoint  torus block can be built by acting with a number of OPE operators on the torus block in the necklace channel with less points.  Section \bref{sec:ceq} considers the  global torus blocks. In Section \bref{sec:global} we formulate the global $n$-point torus blocks and further  in Section \bref{sec:arbitrary} we describe in more detail how to obtain torus blocks in general channels  by acting with $sl(2, \mathbb{R})$ OPE operators on the necklace global block.  In Section \bref{sec:casimir} we present the system of the Casimir equations for $n$-point global torus blocks in the necklace channel and discuss their general properties. In particular, by  change of coordinates,  the torus Casimir equations are conveniently expressed in terms of the counting operators.   In Section \bref{sec:PLimit} we obtain the Casimir equations in the plane limit $q \rightarrow 0$.  In Section \bref{sec:any} we describe the general scheme of building Casimir equations for global torus blocks in arbitrary channels. Section \bref{sec:conclusion} summarizes our results and discusses future perspectives. In Appendix \bref{app:2pt_der} we explicitly derive  the 2-point  Casimir equations and consider the plane limit as well as the reduction to the 1-point case. In Appendix \bref{app:block} we list explicit expressions for some lower-point global torus blocks.

%%%%%%%%%%%%%%%%%%%%%%%%%%%%%%%%%%%%%%%%%
\section{Torus conformal blocks}
\label{sec:def}
%%%%%%%%%%%%%%%%%%%%%%%%%%%%%%%%%%%%%%%%%

A two-dimensional torus $\torus$ can be defined as $\torus \simeq \mathbb{R}^2/{\mathbb{L}}$, where $\mathbb{R}^2$ is a two-dimensional Euclidean plane and  $\mathbb{L} = \mathbb{Z}+\tau \mathbb{Z}$ is a lattice generated by two fundamental periods $1, \tau \in \mathbb{R}^2$. The modular invariant characterization of $\torus$ is achieved in terms of the modulus $\tau$ taking values in the fundamental domain $\subset \mathbb{H}$. In local coordinates $(w, \bar w) \in \mathbb{R}^2$ the torus is therefore realized through the identifications $w \sim w+ 1$ and $w \sim w+ \tau$. Equivalently, introducing  coordinates  $z  =e^{2\pi i w}$, the identification reads $z \sim qz$, where  $q = e^{2\pi i \tau}$, $q \bar q\leq1$, is the modular parameter. In the latter case, $\torus$ can be viewed as an annulus with identified and rotated boundaries. The above quotient parameterization  is extremely useful when considering the torus CFT$_2$. It follows that the basic techniques of the plane \cft can be directly used provided that the double periodicity conditions are imposed on local operators in $w$-coordinates or the scaling conditions in $z$-coordinates.

Local conformal symmetries in the torus \cft are governed  by Virasoro algebra $Vir\oplus \bar Vir$ spanned by  $L_m, \bar L_n,  m,n \in \mathbb{Z}$, the central charge is $c$: $[L_m, L_n] = (m-n)L_{m+n}+(c/12)m(m^2-1)\delta_{m+n,0}$ plus the same  anti-chiral commutation relations.  The only global symmetry is  translational, $u(1)\oplus u(1)\subset Vir \oplus \bar Vir$, which is generated by $L_0$ and $\bar L_0$. The representation space is $\cV_{h,c} \otimes \bar{\cV}_{\bar{h},c}$. Here, the chiral Verma module $\cV_{h,c}$ is spanned by basis descendants
$|M,h\rangle =  L_{-k_1}^{i_1} \cdots L_{-k_n}^{i_n} |h\rangle$,  where $|h\rangle$ is a highest-weight vector of conformal dimension $h$ and $|M| = i_1 k_1 + \cdots + i_n k_n$ with $1 \leq k_1 \leq \cdots \leq k_n$ denotes a level. The Gram matrix in $\cV_{h,c}$ is defined by the inner product of elements on each level by  $B_{M|N} = \langle h,M|N,h\rangle$. The same  relations hold for the anti-chiral Verma module $\bar{\cV}_{\bar{h},c}$. The general  vector is denoted by $|M,\bar{M}, h,\bar{h}\rangle \in \cV_{h,c} \otimes \bar{\cV}_{\bar{h},c}$.

%%%%%%%%%%%%%%%%%%%%%%%%%%%%%%%%%%%%%%%%%
\subsection{Correlation functions}
\label{sec:correlation}
%%%%%%%%%%%%%%%%%%%%%%%%%%%%%%%%%%%%%%%%%

The Hilbert space of states is  $\mathcal{H} = \bigoplus_{D} \cV_{h,c} \otimes \bar{\cV}_{\bh,c}\,$, where conformal dimensions $h, \bh \in D$ and the domain $D$ depends on a particular theory.
The partition function on $\torus$ is given by the following trace over $\cH$:
\be
\label{partition_Z}
Z =  \operatorname{Tr}_\cH\big(q^{L_{0}-\frac{c}{24}} \bar{q}^{\bar{L}_{0}-\frac{c}{24}}\big)\,.
\ee
Then, torus \cft (normalized) correlation functions of $n$ primary local operators $\tilde{\cO}_{h_i, \bar h_i}(w_i, \bar{w}_i)\\\equiv \tilde{\cO}_{i}(w_i, \bar{w}_i)$ with conformal dimensions $(h_i, \bh_i)$ are defined as
\begin{equation}
\label{gen_def}
\big\langle \tilde{\cO}_{1}(w_{1}, \bar w_{1}) \cdots \tilde{\cO}_{n}(w_{n}, \bar{w}_{n})\big\rangle=\frac{(q \bar{q})^{-\frac{c}{24}}}{Z} \operatorname{Tr}_\cH\Big(q^{L_{0}} \bar{q}^{\bar{L}_{0}} \tilde{\cO}_{1}\left(w_{1}, \bar{w}_{1}\right) \ldots \tilde{\cO}_{n}\left(w_{n}, \bar{w}_{n}\right)\Big)\,.
\end{equation}
For further analysis  of the conformal block decomposition  the prefactor in front of the trace  is inessential and can be dropped out.  Furthermore, it is more convenient to use $z$-coordinates so  the correlation functions \eqref{gen_def} take the form
\be
\label{FromwToz}
\begin{split}
\big\langle\cO_{1}(z_{1},\bar{z}_1) \cdots \cO_{n}(z_{n},\bar{z}_n)\big\rangle
=& (2\pi i)^{-\sum_{i=1}^n h_i}\,(-2\pi i)^{-\sum_{i=1}^n \bar{h}_i}\, \Big[\prod_{i=1}^n z_i^{-h_i}\Big]\, \Big[\prod_{i=1}^n \bar{z}_i^{-\bar{h}_i}\Big]\,\\
&\big\langle\tilde{\cO}_{1}(w_{1}(z_1),\bar{w}_1(\bar{z}_1)) \cdots \tilde{\cO}_{n}(w_{n}(z_n),\bar{w}_n(\bar{z}_n))\big\rangle\,,
\end{split}
\ee
where $\cO_{i}(z_{i},\bar{z}_i)$ are related to $\tilde{\cO}_{i}(w_i,\bar{w}_i )$ by the conformal map, so that the difference between two parameterizations is given by the power law prefactor.

The Virasoro symmetry acts on  primary operators  $\cO_{h,\bar{h}}(z,\bar{z})$ in the standard fashion as\footnote{Note that $\cL_n$ form the algebra with opposite sign  structure constants, i.e. $[\cL_m,\cL_n]=-(m-n)\cL_{m+n}$.}
\be
\label{diff_L}
[L_m,\cO_{h,\bar{h}}(z,\bar{z})] = \cL_m \cO_{h,\bar{h}}(z,\bar{z})\,, \qquad   \cL_m = z^m\left[z\partial_z +h(m+1)\right]\,,
\ee
plus the analogous relations for $\bar L_m$. It constrains  the torus correlation functions   \eqref{gen_def} to satisfy the Ward identities which generalize those on the plane and reproduce the standard OPE with the stress tensor \cite{Eguchi:1986sb}. The global Ward identity which is associated with the  (chiral) translational $u(1)$ symmetry reads 
\be
\label{ward_id}
\sum_{i=1}^n \cL_0^{(i)}\, \big\langle\cO_1(z_1,\bar{z}_1)\ldots\cO_i(z_i,\bar{z}_i)\ldots \cO_n(z_n, \bar z_n)\big\rangle = 0\,.
\ee
In particular, 1-point functions are obviously coordinate independent though still have non-trivial modular covariance property \cite{Felder:1989vx} that stems from the modular invariance of the partition function \eqref{partition_Z}.\footnote{See also recent developments in e.g. \cite{Kraus:2016nwo,Brehm:2018ipf}.} In the $n$-point case, the translational invariance fixes the form of  correlation functions up to a {\it leg factor} $\mathbb{L}$ thereby making it an arbitrary function $\mathbb{V}$ of $u(1)$--invariant coordinate combinations $x_i=x_i(z_{i}/z_{i-1})$ and conjugated $\bar x_i$, $i = 2,..., n$. Namely,
\begin{equation}
\label{DecompOfCorr}
\big\langle\cO_1(z_1,\bar{z}_1)\ldots\cO_n(z_n,\bar{z}_n)\big\rangle = \mathbb{L}(\bz) \bar{\mathbb{L}}(\bar{\bz}) \mathbb{V}(q,\bar{q},\bx,\bar{\bx})\,,
\qquad
\mathbb{L}(\bz) = \prod_{i=1}^n z_i^{-\dl_i}\,,
\end{equation}
where $\bz = \{z_1, \ldots, z_n \}$,  $\bx = \{x_2,\ldots x_{n}\}$, and $\bar{\mathbb{L}}(\bar{\bz})$ is obtained from $\mathbb{L}(\bz)$ by substituting  $z\to \bar{z}$ and $h_i \to \bar{h}_i$. The above  leg factor is somewhat canonical but can be redefined by multiplying by a function of $\bx$. Note that in $w$--coordinates this leg factor cancels the same prefactor in \eqref{FromwToz} and the correlation function  \eqref{gen_def} becomes a function of $u(1)$--invariant combinations $w_{i+1} - w_i$.

%%%%%%%%%%%%%%%%%%%%%%%%%%%%%%%%%%%%%%%%%
\subsection{Conformal blocks and  diagrams}
\label{sec:block_diag}
%%%%%%%%%%%%%%%%%%%%%%%%%%%%%%%%%%%%%%%%%

The conformal block decomposition  of  torus correlation functions in a given channel $\mathfrak{Ch}$  reads
\begin{equation}
\label{BlocksExpansion}
\big\langle\cO_1(z_1,\bar{z}_1)\ldots\cO_n(z_n,\bar{z}_n)\big\rangle
= \sum_{\td,\bar{\td}\in D}\mathfrak{Ch}_{h,\tilde{h}}(C)\;
\cF^{\,^{\mathfrak{Ch}}}_{h,\td,c}(q,\bz) \bar{\cF}^{\,^{\mathfrak{Ch}}}_{\bar{h},\bar{\td},c}(\bar{q},\bar{\bz}) \,,
\end{equation}
where  $\cF^{\,^{\mathfrak{Ch}}}_{\dl,\td,c}(q,\bz)$ ($\bar{\cF}^{\,^{\mathfrak{Ch}}}_{\bar{h},\bar{\td},c}(\bar{q},\bar{\bz})$) is the holomorphic (antiholomorphic) conformal block,  coefficients $\mathfrak{Ch}_{h,\tilde{h}}(C)$ are $n$-th order monomials  of structure constants\footnote{\label{fn:str} The structure constants are defined through the  matrix elements as $C_{ijk}  = z_j^{h_k+h_j-h_i} \bar{z}_j^{\bar{h}_k+\bar{h}_j-\bar{h}_i}\langle  h_i,\bar{h}_i| \cO_j(z_j,\bar{z}_j) |h_k,\bar{h}_k \rangle$.} $C_{ijk}$ which form is completely fixed by the choice of $\mathfrak{Ch}$. Here and below, $ h, \bar h \equiv \{{h}_i, \bar{h}_i\in D|\, i = 1,..., n\}$ and $\td, \bar{\td} \equiv \{{\td}_i, \bar{\td}_i\in D|\, i = 1,..., n\}$ are  external and intermediate dimensions, respectively.

The global Ward identity \eqref{ward_id} fixes  a given  $n$-point (holomorphic) torus block to be a  product of the leg factor and a {\it bare} conformal block,
\begin{equation}
\label{leg}
\cF^{\,^{\mathfrak{Ch}}}_{h,\td,c}(q,\bz) = \mathbb{L}(\bz)\; \cV^{\,^{\mathfrak{Ch}}}_{h,\td,c}(q,\bx)\,,
\end{equation}
cf. \eqref{DecompOfCorr}.  The leg factor $\mathbb{L}(\bz)$ here is entirely responsible for the behaviour of the conformal block $\cF^{\,^{\mathfrak{Ch}}}_{h,\td,c}$ with respect to the global conformal transformations.  Thus, the bare conformal block $\cV^{\,^{\mathfrak{Ch}}}_{h,\td,c}$ is $u(1)$--invariant. Moreover, by choosing appropriate variables $(q, \bx) \mapsto {\bm t} \equiv (t_1,...,t_n)$ along with some new leg factor the bare conformal block can be defined to satisfy  the boundary condition  $\cV^{\,^{\mathfrak{Ch}}}_{h,\td,c}(\bm t) = 1$ at $\bm t = 0$.

Evaluating the trace in \eqref{gen_def} yields 
\begin{equation}
\label{TorusFuncThroughPlaneElements}
\begin{split}
\big\langle\cO_1(z_1,\bar{z}_1)\ldots\cO_n(z_n,\bar{z}_n)\big\rangle = \sum_{\tilde h_1,\bar{\td}_1\in D}\;\;\;\sum_{m,\bar m=0}^\infty q^{\tilde h_1+m} \bar{q}^{\bar{\td}_1+\bar m}
 \sum_{m=|M|=|N|} \sum_{\bar m=|\bar{M}|=|\bar{N}|} \\\,B_1^{M|N} B_1^{\bar{M}|\bar{N}}\big\langle\tilde h_1,\bar{\td}_1,M,\bar{M}|\cO_1(z_1,\bar{z}_1)\ldots\cO_n(z_n,\bar{z}_n)|N,\bar{N},\tilde h_1,\bar{\td}_1\big\rangle\,,
\end{split}
\end{equation}
where  $\tilde h_1$ and $\bar{\td}_1$ stand for  the first intermediate conformal dimensions,  $B_1^{M|N}$ and $B_1^{\bar{M}|\bar{N}}$ are elements of the inverse Gram matrices in the Verma modules $\cV_{\tilde h_1,c}$  and $\bar{\cV}_{\bar{\td}_1,c}$.   Thus, $n$-point torus correlation functions are expressed in terms of $(n+2)$-point matrix elements.\footnote{Of course, a straightforward calculation of the torus matrix elements rapidly gets complicated. One can use different approximation schemes when the conformal dimensions and/or the central charge are large. For instance, in the large-$c$ regime one can use the monodromy method, see e.g. \cite{Menotti:2018jsy,Gerbershagen:2021yma,Piatek:2013ifa}, which is also relevant in the context of \adscft.} This formula makes manifest the plane limit   $\im \tau \to \infty$ or $q\to0$, where the $n$-point torus correlation function  gives the plane $(n+2)$-correlation function because the contribution of descendants is suppressed and the leading order defines the product of $n$ primary operators sandwiched between two primary states of weights $\td_1$.  Geometrically, the plane limit can be  seen as going to the infinitely thin and long torus which can effectively be understood as an infinite cylinder conformally mapped onto the plane.  In  Section \bref{sec:PLimit} we will discuss the plane limit of  conformal blocks in the necklace channel.

From now on, we will focus on holomorphic conformal blocks only. The conformal blocks are built of matrix elements so that all calculations are reduced to operations with  chiral conformal algebra generators acting in  chiral Verma modules. It is legitimate because the representation of the full Virasoro algebra is factorized into (anti)chiral subspaces, hence, the action of the conformal generators is also factorized, cf. \eqref{diff_L}. In particular, the trace in \eqref{TorusFuncThroughPlaneElements} is
arranged in such a way that one could first take a partial trace over (anti)chiral subspace $\subset \cH$. Therefore, for convenience, in what follows we suppress any antichiral dependence thereby assuming that all local operators are holomorphic.

The relation \eqref{TorusFuncThroughPlaneElements} is the first step in defining a channel $\mathfrak{Ch}$ of the conformal block decomposition  \eqref{BlocksExpansion} as it singles out matrix elements with some $\tilde h_1 \in D$.  The  channel $\mathfrak{Ch}$ can be further specified  by doing   OPE  and/or inserting projectors in the matrix elements   \eqref{TorusFuncThroughPlaneElements}
\be
\label{1st_inter}
\sum_{m=0}^\infty \sum_{m=|M|=|N|} q^{\tilde h_1+m} \,B_1^{M|N}\big\langle\tilde h_1,M|\cO_1(z_1)\ldots\cO_n(z_n)|N,\tilde h_1\big\rangle\,.
\ee
Namely, any two primary operators $\cO_{h_i}$ and $\cO_{h_j}$ in \eqref{TorusFuncThroughPlaneElements} can be fused by means of the OPE  $\cO_{h_i} \cO_{h_j} \subset [\cO_{\td}]$ to produce a secondary operator belonging to the conformal family of dimension $\td \in D$ either one can insert  between them a projector on the Verma module $\cV_{\td,c}$. A particular combination of these operations defines a channel $\mathfrak{Ch}$ and, hence,  the corresponding conformal block. Each OPE introduces  one intermediate dimension $\td$ and reduces the number of primary operators by one so that after $(n-k)$ OPEs there are $k$ primary operators left. Then,  between the remaining operators one inserts $(k-1)$ projectors with other intermediate dimensions. Taking into account that one intermediate  dimension $\td_1$ comes from the trace  \eqref{TorusFuncThroughPlaneElements}, there are  $(n-k)+(k-1)+1 = n$ intermediate dimensions characterising a particular channel $\mathfrak{Ch}$, cf. \eqref{BlocksExpansion}.

\begin{figure}%[H]
\begin{minipage}[h]{0.40\linewidth}
\center{\includegraphics[width=1\linewidth]{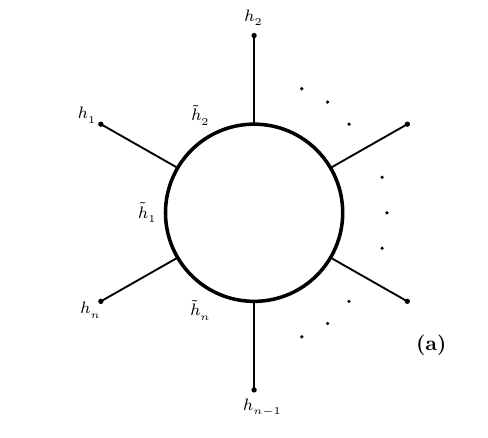}}
\end{minipage}
\begin{minipage}[h]{0.45\linewidth}
\center{\includegraphics[width=1\linewidth]{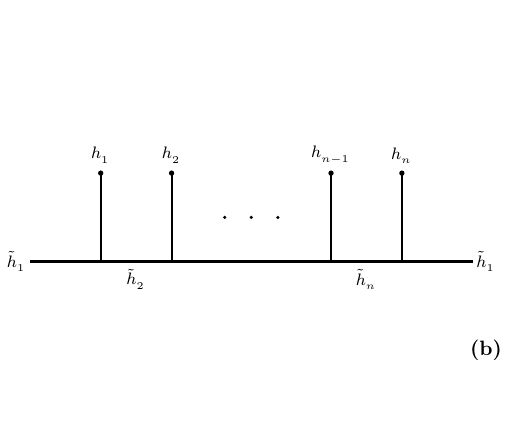}}
\end{minipage}
\caption{The $n$-point torus block in the necklace channel {\bf (a)}. It can be thought of as the $(n+2)$-point block in the comb channel of the plane \cft with identified endpoints {\bf (b)}. Note that $\td_1$ on the necklace diagram is always between $h_1$ and $h_n$.}
\label{Necklace}
\end{figure}

\begin{figure}%[H]
\centering
\includegraphics[width=0.25\linewidth]{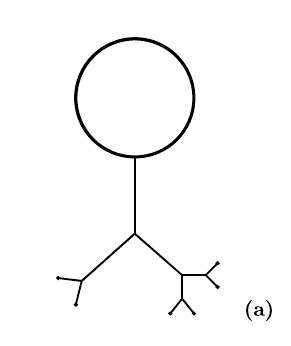}
\qquad
\qquad
\includegraphics[width=0.30\linewidth]{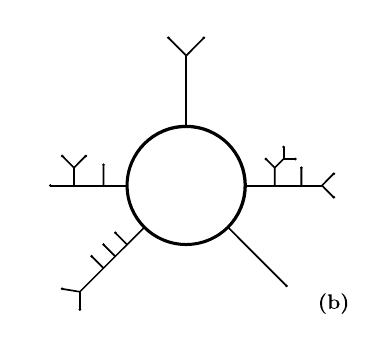}

\caption{ {\bf (a)} A particular OPE channel for 6-point torus block. {\bf (b)} Arbitrary torus channel $\mathfrak{Ch}$.}
\label{f1}
\end{figure}

\paragraph{Necklace channel.} Depending on whether only OPEs or projectors were used there are two distinguished  channels: the OPE channel and the projection channel. Let us consider the projection channel which we will refer to as the necklace channel following the form of its diagram shown in fig. \bref{Necklace}.\footnote{The OPE channel is when only one operator along with its conformal family is connected to the necklace which therefore has only one leg \cite{Kraus:2017ezw}, see fig. \bref{f1} {\bf (a)}.} For blocks of this type we omit the superscript $\mathfrak{Ch}$, for the sake of simplicity. The corresponding block function can be formally defined by considering holomorphic matrix element \eqref{1st_inter} and inserting the projectors between each pair of primary operators:  
\begin{equation}
\label{neck_block_n}
\cF_{h,\td,c}(q,\bz)
= \left[\prod_{i=1}^n\frac{z_i^{h_i+\tilde h_{i+1}-\tilde h_i}}{ \langle  \tilde h_i| \cO_i(z_i) |\tilde h_{i+1} \rangle}\right] \Tr_{\td_1}\left[q^{L_0}\cO_1(z_1) \mathbb{P}_2 \cO_2(z_2)\ldots \mathbb{P}_n \cO_n(z_n)\right] \,,
\end{equation}
where, in the prefactor, we identify $\td_{n+1}\equiv \td_1$, the partial trace $\Tr_{\td_1}$ is taken over chiral part of the Hilbert space, and the projectors are given by
\begin{equation}
\label{VirProjector}
\mathbb{P}_i = \sum_{m_i=0}^\infty \;\sum_{m_i=|M|=|N|}B_i^{M|N}|N,\td_i\rangle\langle\td_i,M|,\qquad
\mathbb{P}_i^2 = \mathbb{P}_i\,,
\qquad
\sum_{\td_i\in D}\mathbb{P}_i = 1,
\end{equation}
with $B_i^{M|N}$ being  inverse Gram matrices in $\cV_{\td_i,c}$. Then,   the  bare torus block in the necklace channel is given by 
\begin{equation}
\label{bare_block}
\cV(q,\bx) = \left[\prod_{i=1}^n \frac{z_i^{2h_i+\tilde h_{i+1}-\tilde h_i}}{\langle  \tilde h_i| \cO_i(z_i) |\tilde h_{i+1} \rangle}\right]\Tr_{\td_1}[q^{L_0}\cO_1(z_1) \mathbb{P}_2 \cO_2(z_2)\ldots \mathbb{P}_n \cO_n(z_n)]\,.
\end{equation}

Restoring the full (anti)holomorphic dependence and coming back to the original correlation function \eqref{gen_def} one obtains the conformal block decomposition in the  necklace channel by inserting full projectors $\mathbb{P}_i\bar{\mathbb{P}}_i$ so that the matrix element \eqref{TorusFuncThroughPlaneElements} gets represented as a product of 3-point functions of primary and secondary operators. Stripping off descendants both in the (anti)chiral sectors by commuting $L_m$ and $\bar L_n$ with primary operators $\cO(z_i, \bar z_i)$ one finally obtains 
\be
\label{add1}
\big\langle\cO_1(z_1,\bar{z}_1)\ldots\cO_n(z_n,\bar{z}_n)\big\rangle 
= \hspace{-2mm} \sum_{\tilde h,\bar{\td}\in D}
\cF_{h,\td,c}(q,\bz) \bar{\cF}_{\bar{h},\bar{\td},c}(\bar{q},\bar{\bz}) 
\Bigg(\prod_{i=1}^n \frac{\langle  \tilde h_i, \bar{\tilde h}_i| \cO_i(z_i, \bar z_i) |\tilde h_{i+1}, \bar{\tilde h}_{i+1} \rangle}{z_i^{h_i+\tilde h_{i+1}-\tilde h_i}\, \bar{z}_i^{\bar{h}_i+\bar{\tilde{h}}_{i+1}-\bar{\tilde h}_i}}\Bigg),
\ee
where the (anti)holomorphic blocks come up as the result of the straightforward  calculation while  the remaining factor is the product of 3-point functions of primary operators which are proportional to  the structure constants (see the footnote \bref{fn:str}). Thus, the structure constant prefactor  $\mathfrak{Ch}_{h,\tilde{h}}(C)$ in the conformal block decomposition \eqref{BlocksExpansion} reads
\be
\mathfrak{Ch}_{h,\tilde{h}}(C) = \prod_{i=1}^n C_{\td_i,h_i,\td_{i+1}}\,.
\ee

The tricky point here is that if we are interested in the holomorphic block only and not in the full conformal block decomposition \eqref{add1}, we can independently  define the conformal block  as a holomorphic function resulting from evaluating  the formal holomorphic matrix element on the right-hand side in \eqref{neck_block_n}. The general (anti)holomorphic factorization in the full \cft  guarantees that the representation \eqref{neck_block_n} can always be used once the antiholomorphic dependence is suppressed.

\paragraph{General channel.} Using the quotient description of $\mathbb{T}^2$, the necklace channel can be viewed as  obtained  from the comb channel  on the plane (fig. \bref{Necklace}) by gluing together  the endpoints $z=\infty$ and $z=0$ (first and last operators) and summing up over all descendants that produces the trace $\Tr_{\td}$. Such a gluing can be done for any conformal block diagram on $\mathbb{R}^2$ yielding a conformal block diagram on $\mathbb{T}^2$ which is a necklace with $k \leq n$ legs   to which $k$ graphs of arbitrary topology\footnote{Here, by a topology we mean a plane graph obtained by gluing single node trivalent trees  in a particular order without making loops. See \cite{Fortin:2020zxw,Fortin:2020yjz,Fortin:2020bfq} for discussion of the plane \cft conformal blocks described by diagrams of arbitrary topology.} are attached, see fig. \bref{f1}. Specific plane graphs correspond to a particular ordering  of  OPEs in the torus correlation function.

In what follows, we denote the channel topology described above as $\mathfrak{Ch}$ = $(k-1, \mathfrak{ch})$, where the number of inserted projectors comes first and the second symbol $\mathfrak{ch}$ stands for a certain topology of trivalent trees obtained by doing $(n-k)$ ordered OPEs. It is clear that the number of projectors producing  a loop in  respective torus block diagrams is a natural characterization of all possible  diagrams by a number of intermediate lines forming a loop, see fig. \bref{f1}. On the other hand, $\mathfrak{ch}$ embraces all possible OPEs with intermediate legs attached to the loop.

A torus conformal block of an arbitrary channel topology can be generated from the  torus conformal block with less points in the necklace channel.\footnote{The comb channel in the plane \cft plays the analogous role.} One can consider a $n$-point block, which is obtained from the $n$-point correlation function \eqref{TorusFuncThroughPlaneElements} by inserting $(k-1)$ projectors where $k = 1, 2,\ldots$ and doing $(n-k)$ OPEs. On the other hand, each OPE contains Virasoro generators which can be represented  as differential operators   acting on the correlation function \cite{Belavin:1984vu}.  Then, the resulting OPE coefficients can be transformed into some differential operator $\hat B^{\mathfrak{ch}}_{h,\td,c}$ acting on a $k$-point torus block $\cF_{h,\td,c}(q,z'_1,\ldots,z'_k)$ in the necklace channel:
\be
\label{gc}
\cF^{\,(k-1, \mathfrak{ch})}_{h,\td,c}(q,z_1,\ldots,z_n) = \hat B^{\mathfrak{ch}}_{h,\td,c} (z_1, \ldots, z_n, \partial_{z'}) \cF_{h,\td,c}(q,z'_1,\ldots,z'_k) \,,
\ee
where the set  $(z'_1,...,z'_k)$ denotes coordinates left after doing  all OPEs. The explicit form of $\hat B^{\mathfrak{ch}}_{h,\td,c}$ is directly determined by the topology of a channel $\mathfrak{ch}$ which for the general Virasoro conformal blocks can be  quite complicated. However, in the $sl(2, \mathbb{R})$ case, these operators take a simpler form, see Section \ref{sec:arbitrary}. Notice that the block on the left-hand side of \eqref{gc}  automatically satisfies the Ward identity provided that the block on the right-hand side  does.

To conclude, it is instructive to compare the mixed  use of the  projector/OPE technique in torus and plane correlation functions. For instance, for  the 4-point correlation function $\langle \cO_1(z_1) \cO_2(z_2)\cO_{3}(z_{3})\cO_4(z_4)\rangle$ on the plane this formally  yields several options. There are three standard OPE channels ($s,t,u$) obtained by doing OPEs between pairs of operators, depending on how the points $(z_1=0, z_2=1, z_3=z, z_4=\infty)$ are ordered. However, one can see that the only non-trivial way to obtain the projection  channel is to insert the projector between pairs $\cO_1(z_1) \cO_2(z_2)$ and $\cO_{3}(z_{3})\cO_4(z_4)$.  On the other hand, it is straightforward to show that the projection channel coincides with the $s$-channel (see e.g.  \cite{Simmons-Duffin:2016gjk}). Thus, all channels on the plane are the OPE channels. As mentioned above, $n$-point torus correlation functions  can be thought of as identifying points $z_1=0 \sim z_{n+2} = \infty$ and taking the trace over descendants in  $(n+2)$-point plane correlation functions. It makes one to  distinguish  between   projectors  and  OPEs  as operations leading to (topologically) different channels. For example, the plane $t$-channel directly yields the torus OPE channel, while the plane  $s,u$-channels can yield only one torus channel which is the projection  (necklace) channel. In the latter case the equivalence between the projector and the OPE on the plane  is broken in favour of the projector on the torus. This explains why $n$-point torus correlators  can be expanded in more (topologically inequivalent) channels than $(n+2)$-point and, hence, $n$-point plane correlators.

%%%%%%%%%%%%%%%%%%%%%%%%%%%%%%%%%%%%%%%%%
\section{Casimir equations for $n$-point global torus blocks}
\label{sec:ceq}
%%%%%%%%%%%%%%%%%%%%%%%%%%%%%%%%%%%%%%%%%

Let us  consider the large-$c$ limit. It can be shown that the Virasoro algebra and its representations can enjoy at least two consistent Inonu-Wigner contractions in $1/c$, one of which yields global conformal blocks and another  one yields light conformal blocks which in the torus \cft are different but related functions \cite{Alkalaev:2016fok,Cho:2017oxl}. In this section we focus on global conformal blocks $\cF^{\,^{\mathfrak{Ch}}}_{\dl,\td}(q,z)$ which are associated with the finite-dimensional subalgebra  $sl(2,\mathbb{R})\subset Vir$.

%%%%%%%%%%%%%%%%%%%%%%%%%%%%%%%%%%%%%%%%%
\subsection{Global torus  blocks in the necklace channel}
\label{sec:global}
%%%%%%%%%%%%%%%%%%%%%%%%%%%%%%%%%%%%%%%%%

The $sl(2,\mathbb{R})$ Verma modules $V_h$ are spanned by basis vectors obtained  from the highest weight vector $|\dl\rangle$ as $\{|m,\dl\rangle = L_{-1}^m|\dl\rangle, \; L_0|\dl\rangle = h|\dl\rangle\}$, where $L_{0, \pm 1}$ are $sl(2, \mathbb{R})$ generators. Thus,  each level contains just one basis element. The conformal dimensions are assumed to be arbitrary real numbers, $h \in \mathbb{R}$. The inverse Gram matrix is diagonal, $B^{M|N} := B^{m} = (m!(2h)_m)^{-1}$, where $(a)_m = a(a+1)\ldots(a+m-1)$ is the Pochhammer  symbol. Then, the projector \eqref{VirProjector} onto intermediate Verma modules $V_{\td_i}$ takes the form
\begin{equation}
\label{sl2Projector}
\mathbb{P}_i = \sum_{m=0}^\infty \frac{|m,\td_i\rangle \langle \td_i,m|}{m!(2\td_i)_m}\,.
\end{equation}

All the constructions discussed in the previous section for the Virasoro conformal blocks directly apply to the $sl(2, \mathbb{R})$ global blocks. In particular, substituting the projectors \eqref{sl2Projector} into the definition \eqref{neck_block_n} yields the following $n$-point torus block function in the necklace channel \cite{Alkalaev:2017bzx}
\be
\label{nglobs}
\cV_{h,\td}(x) = \Big(\prod_{j=1}^n x_j^{\td_{j}}\Big)\sum_{s_1,...,s_{n}=0}^\infty \;\;\prod_{m=1}^n \frac{\tau_{s_m,s_{m+1}}(\td_{m}, h_m, \td_{m+1})}{s_m!\,(2\td_m)_{s_m} }\,  x_m^{s_m}\,,
\ee
where $x_j=z_{j}/z_{j-1}$, $j=2,\ldots,n$, $x_1 = q/(x_2\ldots x_{n})$ with identifications $\td_{n+1}=\td_1$, $s_{n+1}=s_1$ and $\tau$-coefficients are given by \cite{Alkalaev:2015fbw}
\begin{equation}\label{TauElement}
\tau_{m,n}(a,b,c) = m!n!\sum_{p=0}^{min[m,n]}\frac{(2c+n-1)^{(p)}(c+b-a)_{n-p}(a+b-c+p-n)_{m-p}}{p!(m-p)!(n-p)!}\,.
\end{equation}
Note that it would be useful to find the block function \eqref{nglobs} in the form similar to that one in \cite{Rosenhaus:2018zqn}, where $n$-point global block of the plane \cft was represented as a special function generalizing the Appel series (see \eqref{CombBlock} below). Hopefully, the Casimir approach developed in the next section will provide more insight into  the formulation  of the global torus block functions (see discussion in the end of Section \bref{sec:PLimit}).

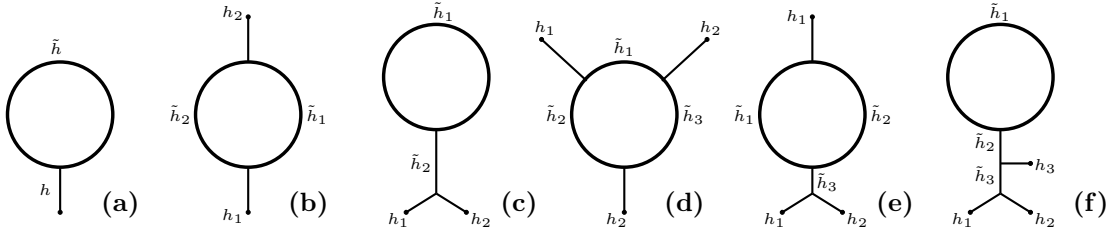
\begin{figure}%[H]
\centering
\begin{tikzpicture}[scale=1.0]
 \tkzDefPoint(0,0){X}
  \tkzDefPoint(0.7,0){X1}

  \tkzDefPoint(2.5,0){A}
   \tkzDefPoint(3.2,0){A1}

   \tkzDefPoint(5,0.5){B}
    \tkzDefPoint(5.7,0.5){B1}

      \tkzDefPoint(7.5,0){C}
      \tkzDefPoint(8.2,0){C1}

        \tkzDefPoint(10,0){D}
         \tkzDefPoint(10.7,0){D1}

          \tkzDefPoint(12.5,0.5){E}
             \tkzDefPoint(13.2,0.5){E1}

             \tkzDrawCircle[line width  = 1.3](X1,X)
                 \tkzDrawCircle[line width  = 1.3](A1,A)
                 \tkzDrawCircle[line width  = 1.3](B1,B)
                 \tkzDrawCircle[line width  = 1.3](C1,C)
                  \tkzDrawCircle[line width  = 1.3](D1,D)
                  \tkzDrawCircle[line width  = 1.3](E1,E)

            % a)
        \draw[color = black, line width  = 0.8] (0.7,-0.7) -- (0.7, -1.3);

        %b)
        \draw[color = black, line width  = 0.8] (3.2,-0.7) -- (3.2, -1.3);
        \draw[color = black, line width  = 0.8] (3.2,0.7) -- (3.2, 1.3);

        %c)
        \draw[color = black, line width  = 0.8] (5.7,-0.2) -- (5.7, -1.05);
        \draw[color = black, line width  = 0.8]  (5.7, -1.05) -- (5.3, -1.3);
         \draw[color = black, line width  = 0.8]  (5.7, -1.05) -- (6.1, -1.3);

        % d)

        \draw[color = black, line width  = 0.8] (8.2,-0.7) -- (8.2, -1.3);
         \draw[color = black, line width  = 0.8] (8.7, 0.45) -- (9.3, 1.0);
           \draw[color = black, line width  = 0.8] (7.7, 0.45) -- (7.1, 1.0);

         % e)
         \draw[color = black, line width  = 0.8] (10.7,0.7) -- (10.7, 1.3);
    \draw[color = black, line width  = 0.8] (10.7,-0.7) -- (10.7, -1.05);
 \draw[color = black, line width  = 0.8]  (10.7, -1.05) -- (11.1, -1.3);
         \draw[color = black, line width  = 0.8]  (10.7, -1.05) -- (10.3, -1.3);

      %f)

        \draw[color = black, line width  = 0.8] (13.6,-0.65) -- (13.2, -0.65);
        \draw[color = black, line width  = 0.8] (13.2,-0.2) -- (13.2, -1.05);
        \draw[color = black, line width  = 0.8]  (13.2, -1.05) -- (12.8, -1.3);
         \draw[color = black, line width  = 0.8]  (13.2, -1.05) -- (13.6, -1.3);

        \tkzDefPoint(0.7, -1.3){a1}

        \tkzDefPoint(3.2, -1.3){b1}
         \tkzDefPoint(3.2, 1.3){b2}

          \tkzDefPoint(5.3, -1.3){c1}
          \tkzDefPoint(6.1, -1.3){c2}

            \tkzDefPoint(8.2, -1.3){d1}
            \tkzDefPoint(9.3, 1.0){d2}
            \tkzDefPoint(7.1, 1.0){d3}

          \tkzDefPoint(10.3, -1.3){e1}
          \tkzDefPoint(12.8, -1.3){e2}
          \tkzDefPoint(10.7, 1.3){e3}

          \tkzDefPoint(13.6, -1.3){f1}
          \tkzDefPoint(11.1, -1.3){f2}
           \tkzDefPoint (13.6, -0.65){f3}

          \draw   (1.5, -1.2) node {\small \textbf{(a)}};

          \draw   (0.5, -1.0) node {\tiny \textbf{$h$}};

          \draw   (0.65, 0.9) node {\tiny \textbf{$\tilde h$}};

          \draw   (4.0, -1.2) node {\small \textbf{(b)}};

\draw   (3.0, -1.35) node {\tiny $h_{1}$};
\draw   (3.0, 1.35) node {\tiny $h_{2}$};
\draw   (2.3, 0.0) node {\tiny $\tilde{h}_{2}$};
\draw   (4.12, 0.0) node {\tiny $\tilde{h}_{1}$};

            \draw   (6.8, -1.2) node {\small \textbf{(c)}};

\draw   (5.8, 1.37) node {\tiny $\tilde h_1$};
\draw   (5.5, -0.65) node {\tiny $\tilde h_2$};
\draw   (5.2, -1.4) node {\tiny $h_1$};
\draw   (6.3, -1.4) node {\tiny $h_2$};

             \draw   (9.0, -1.2) node {\small \textbf{(d)}};

\draw    (7.15, 1.15)  node {\tiny $h_1$};
\draw    (9.35, 1.15)  node {\tiny $h_2$};
\draw    (8.2, 0.9)  node {\tiny $\tilde h_1$};
\draw    (8.1, -1.4)  node {\tiny $h_2$};
\draw   (7.3, 0.0) node {\tiny $\tilde{h}_{2}$};
\draw   (9.12, 0.0) node {\tiny $\tilde{h}_{3}$};

              \draw   (11.8, -1.2) node {\small \textbf{(e)}};

\draw   (11.3, -1.4) node {\tiny $h_2$};
\draw   (10.2, -1.4) node {\tiny $h_1$};
\draw   (10.9, -0.9) node {\tiny $\tilde{h}_3$};
\draw   (9.8, 0.0) node {\tiny $\tilde{h}_{1}$};
\draw   (11.62, 0.0) node {\tiny $\tilde{h}_{2}$};
\draw   (10.5, 1.3) node {\tiny $h_{1}$};

               \draw   (14.4, -1.2) node {\small \textbf{(f)}};

\draw   (13.8, -1.4) node {\tiny $h_2$};
\draw   (12.7, -1.4) node {\tiny $h_1$};
\draw   (13.8, -0.68) node {\tiny $h_3$};
\draw   (13.0, -0.81) node {\tiny $\tilde{h}_3$};
\draw   (13.0, -0.37) node {\tiny $\tilde{h}_2$};
\draw   (13.2, 1.36) node {\tiny $\tilde{h}_1$};

          \tkzDrawPoints[color=black,fill=black,size=1.5](a1, b1, b2, c1, c2, d1, d2, d3, e1, e2, e3, f1, f2, f3)

      \end{tikzpicture}
\caption{1-point block {\bf (a)}. 2-point torus block in the necklace channel {\bf (b)}, in the OPE channel {\bf (c)}. 3-point torus blocks in the necklace channel {\bf (d)}, in the mixed channel {\bf (e)}, in the OPE channel {\bf (f)}.}
\label{fig_23pt}
\end{figure}

%%%%%%%%%%%%%%%%%%%%%%%%%%%%%%%%%%%%%%%%%%%%%%%%%%%%%%%%%%%%%
\subsection{Global torus  blocks in arbitrary  channels}
\label{sec:arbitrary}
%%%%%%%%%%%%%%%%%%%%%%%%%%%%%%%%%%%%%%%%%%%%%%%%%%%%%%%%%%%%%%

Let us  consider  the general prescription \eqref{gc} in the case of the global blocks.  For the $sl(2, \mathbb{R})$ conformal symmetry, the OPE of two primary operators is given by
\be
\label{OPE_sl}
\cO_i(z_i) \cO_j(z_j) \sim \hat{\phi}_{h,\td} (z_i, z_j, \partial_{z_j}) \cO_{\td}(z_j)\,,
\ee
where $\sim$ means that we omitted the structure constants which are assumed to be non-vanishing for these three operators, and the OPE operator $\hat{\phi}_{h,\td}$ reads (see e.g. \cite{Belavin:1984vu,Blumenhagen,Qualls:2015qjb})
\be
\label{OPE_oper}
\hat{\phi}_{h,\td} (z_i, z_j, \partial_{z_j})= (z_i-z_j)^{\td-h_i-h_j} \sum_{m=0}^\infty \frac{(\td+h_i-h_j)_m}{m!(2\td)_m}\,(z_i-z_j)^m\, \partial_j^m\,.
\ee
Then, the operators $\hat B^{\mathfrak{ch}}_{h,\td}$ \eqref{gc} are built as sequences of the OPE operators $\hat{\phi}_{h,\td}$  ordered according to the particular OPE ordering of  a given  channel $\mathfrak{ch}$.

The above construction  can be illustrated  by  considering  a few examples of low-point blocks shown  in fig. \bref{fig_23pt}. The 1-point block exists in the only channel {\bf (a)}. The first non-trivial example is given by the $2$-point torus block which can be considered in the necklace channel {\bf (b)} and in the OPE channel {\bf (c)}. In the latter case, the  formula \eqref{gc} takes the form
\be
\label{fex}
\cF^{\,(0,\mathfrak{ch}_1)}_{h_1 h_2, \td_1 \td_2} (q, z_1, z_2) = \hat{\phi}_{h_1 h_2, \td_2} (z_1, z_2, \partial_{z_2}) \cF_{\td_2,\td_1}(q,z_2)\,.
\ee
Thus, the $2$-point block  is expressed in terms of the $1$-point block $\cF_{\td_2,\td_1}(q,z_2)$ {\bf (a)}. The next example is the $3$-point block in three different channels {\bf (d)}, {\bf (e)}, {\bf (f)}. In the OPE channel {\bf (f)}, which directly generalizes the OPE channel {\bf (c)} by adding one more OPE operator, one finds
\be
\label{nex}
\cF^{\,(0,\mathfrak{ch}_2)}_{h_1 h_2 h_3,\td_1 \td_2 \td_3}(q,z_1, z_2, z_3) = \hat{\phi}_{h_1 h_2,\td_3} (z_1, z_2, \partial_{z_2}) \hat{\phi}_{\td_3 h_3, \td_2} (z_2, z_3, \partial_{z_3}) \cF_{\td_2,\td_1}(q,z_3) \,.
\ee
In the mixed channel {\bf (e)} obtained by inserting two projectors and doing one OPE one finds
\be
\label{nexx}
\cF^{\,(1, \mathfrak{ch}_1)}_{h_1 h_2 h_3,\td_1 \td_2 \td_3}(q,z_1, z_2, z_3) = \hat{\phi}_{h_1 h_2,\td_3} (z_1, z_2, \partial_{z_2}) \cF_{h_3 \td_3,\td_1 \td_2} (q, z_2, z_3)\,,
\ee
where  $\cF_{h_3 \td_3,\td_1 \td_2} (q, z_2, z_3)$ is a $2$-point block in the necklace channel {\bf (b)}. The 3-point necklace channel {\bf (d)} can be used to build  torus block with 4 points and higher. Explicit expressions of the blocks considered above are collected in Appendix \bref{app:block}.

%%%%%%%%%%%%%%%%%%%%%%%%%%%%%%%%%%%%%%%%%
\subsection{Casimir equations in the necklace channel}
\label{sec:casimir}
%%%%%%%%%%%%%%%%%%%%%%%%%%%%%%%%%%%%%%%%%

Since the conformal families defining exchange channels form $sl(2,\mathbb{R})$ irreducible representations the global  blocks can be viewed as eigenfunctions of the Casimir operators \cite{Dolan:2003hv} which in the present case is given by
\begin{equation}
\label{C2}
\operatorname{C_2} = L_1 L_{-1} - L_0 - L_0^2\;.
\end{equation}
Its eigenvalue on irreducible modules $V_{h}$ equals $-h(h-1)$.

The associated Casimir equations can be  obtained by inserting the Casimir operator \eqref{C2} between operators under the trace  in the torus block function   \eqref{neck_block_n}. Using the  obvious properties $\text{C}_2\mathbb{P}_i = \mathbb{P}_i\text{C}_2 = -\td_i(\td_i-1)\mathbb{P}_i$ and $\tr_h \text{C}_2\hat A = - h(h-1) \tr_h \hat A$ one finds $n$ distinct relations which realize irreducibility of intermediate  conformal families
\be
\label{Cas_eq_n}
\begin{split}
&1)\quad\Tr_{\td_1}[C_2 q^{L_0}\cO_1 \mathbb{P}_2 \cO_2\cdots \mathbb{P}_n \cO_n] = -\td_1 (\td_1 -1)\cF_{h,\td}\,,\\
&\quad \quad\ldots \\
&j)\quad\Tr_{\td_1}[ q^{L_0}\cO_1 \mathbb{P}_2 \cO_2\cdots \mathbb{P}_j C_2 \cO_j \cdots \mathbb{P}_n \cO_n] = -\td_j(\td_j-1)\cF_{h,\td}\,,\\
&\quad\quad \ldots \\
&n)\quad\Tr_{\td_1}[ q^{L_0}\cO_1 \mathbb{P}_2 \cO_2\cdots \mathbb{P}_n C_2\cO_n] = -\td_n(\td_n-1)\cF_{h,\td}\,,
\end{split}
\ee
where $\cF_{h,\td}= \cF_{h,\td}(q, \bz)$ is the $n$-point torus block in the necklace channel.
Note that in this form the torus Casimir equations are not immediately seen as the eigenvalue problem for the conformal block functions as in the plane \cft case \cite{Dolan:2003hv,Alkalaev:2015fbw,Rosenhaus:2018zqn}.

In Appendix \bref{app:2pt_der} we explicitly demonstrate how to transform the above system to the eigenvalue equations in the 2-point case. Generalizing that procedure to the $n$-point case we find that the global torus block in the necklace  channel is subjected to the following  Casimir equation system  ($j = 2,\ldots, n$):
\be
\label{Casn1}
\ba{l}	
\dps
\bigg[-q^2\partial_q^2 + \frac{2q}{1-q}q\partial_q -\frac{q}{(1-q)^2}\sum_{i=1}^n\cL_{-1}^{(i)}\sum_{k=1}^n\cL_{1}^{(k)}\bigg]\cF_{h,\td} = -\td_1 (\td_1 -1)\cF_{h,\td}\,,
\vspace{4mm}
\\
\dps
\bigg[-q^2\partial_q^2 + \frac{2q}{1-q}q\partial_q - \frac{1}{(1-q)^2}\bigg(\sum_{k=1}^{j-1}\cL_{-1}^{(k)}+q\sum_{k=j}^{n}\cL_{-1}^{(k)} \bigg)\bigg(q \sum_{k=1}^{j-1}\cL_{1}^{(k)}+\sum_{k=j}^{n}\cL_{1}^{(k)}\bigg)
\vspace{2mm}
\\
\dps
\hspace{3mm}+\frac{1+q}{1-q}\sum_{k=j}^n\cL_{0}^{(k)}-\sum_{k=j}^n\cL_{0}^{(k)}\sum_{l=j}^n\cL_{0}^{(l)}-2\sum_{k=j}^n\cL_{0}^{(k)}q\partial_q\bigg]\cF_{h,\td} = -\td_j(\td_j-1)\cF_{h,\td}\,,
\ea
\ee
where $\cL_{m}^{(k)}$  are differential realizations of $sl(2, \mathbb{R})$ generators $L_m$, $m = 0, \pm1$ \eqref{diff_L} at points $z_k$.
A few comments are in order.

First, we note that the first equation in \eqref{Casn1} differs from the other equations due to the way the intermediate dimensions were introduced: the first dimension $\td_1$ arises from the common trace in the definition of the torus function \eqref{TorusFuncThroughPlaneElements}, while the other dimensions $\td_{2,...,n}$ come from the projectors. The equations can be unified by applying  the Ward identity \eqref{ward_id}:
\begin{equation}\label{nWard}
\sum_{i=1}^n \cL_0^{(i)}\cF_{h,\td} = 0\,.
\end{equation}
Then, the first equation in \eqref{Casn1} can be written in the form of the second equation in \eqref{Casn1} by setting $j=1$ and adding zero terms proportional \eqref{nWard}. The resulting  system can be written as ($i=1,...,n$)
\be
\label{Casn1_d}
\cC_2[n,i] \cF_{h,\td}(q, {\bf z}) = -\td_i(\td_i-1)\cF_{h,\td}(q, {\bf z})\,,
\ee
where the differential operators $\cC_2[n,i]$ are read off from the left-hand sides of \eqref{Casn1}.

Second, supplementing the Casimir equation system by the Ward identity one gets $(n+1)$ 2nd order PDEs in the modular parameter $q$ and $n$ insertion points ${\bf z}$ on a single block function $\cF_{h,\td}(q, {\bf z})$. The Ward identity  removes dependence on one of ${\bf z}$-coordinates and, hence, the general solution is parameterized by $2n$ constants which define an asymptotic behaviour of the block function. In general, one expects that different branches would correspond to the original  block in the necklace channel along with the shadow blocks.\footnote{For the shadow blocks in the plane CFT$_d$ see  e.g. \cite{SimmonsDuffin:2012uy}, the shadow formalism was also  used to study CFT$_d$ thermal blocks in \cite{Gobeil:2018fzy}.}

Third, the system \eqref{Casn1} allows reducing a number of points from $n$ to $(n-k)$. In order to do this, one has to take into account that $(n-k)$-point blocks satisfy the Ward identity \eqref{nWard} with $n$ replaced by $(n-k)$ and impose on the $n$-point conformal block $k$ conditions $h_i=0$, $i=n,\ldots,(n-k+1)$, from which it follows that
\begin{equation}
\partial_{z_k}\cF_{h,\td} = 0\,,
\qquad  k=n,\,(n-1),\,\ldots\,, (n-k+1)\,.
\end{equation}
In addition, one has to put $\td_j = \td_1$, $j=n,\ldots,(n-k+1)$.

The Casimir equations \eqref{Casn1} can be considerably simplified by switching to the bare conformal block $\cV(q,\bm x)$ \eqref{leg} and changing variables from $z_i$, $i=1,\ldots,n$ to $x_i=z_{i}/z_{i-1}$, $i=2,\ldots,n$. We find that the Casimir equation system can be cast into the form
\be
\label{CasWithoutLeg}
\cas\, \cV_{h,\td}(q,\bm x) = -\td_j(\td_j-1)\,\cV_{h,\td}(q,\bm x)\,,
\qquad
j = 1,...\,, n\,,
\ee
where the 2nd order differential operators on the left-hand side are given by
\begin{equation}
\label{C2n}
\cas = \frac{1+q}{1-q}\,(N+N_j)-(N+N_j)^2 - \frac{1}{1-q}\bigg(\frac{q}{1-q} \sum_{i,k=1}^n \cD_{i,k}+\sum_{i=1}^{j-1} \sum_{k=j}^n (\cD_{i,k}-q \cD_{k,i})\bigg).
\end{equation}
Here, $N_k \equiv x_k \partial_{x_k}$ and  $N\equiv q\partial_q$ are the counting  operators ($k=1,...,n+1$,  with the convention $N_1\equiv0\equiv N_{n+1}$), and their combinations are
\be
\cD_{i,k} \equiv \Big(\lambda_{i,k}(N_{i}-N_{i+1}-h_i)+\delta_{i,k}\Big)\Big(N_{k}-N_{k+1}+h_k\Big) \,,
\ee
where
\begin{equation}
\label{lambda}
\lambda_{i,k} \equiv \frac{z_k}{z_i} =
\left\{
\begin{split}
&\prod_{l=i+1}^{k}x_l, \quad \mbox{if } k > i,
\\
&1 , \quad \mbox{if } k=i,
\\
&\prod_{l=k+1}^{i}x_l^{-1}, \quad \mbox{if } k < i.
\end{split}
\right.
\end{equation}
Choosing different coordinates one can find other equivalent forms of the Casimir equations. However, the system \eqref{CasWithoutLeg}--\eqref{lambda} is conveniently expressed in terms of the counting operators that could be crucial from the perspective of finding and classifying solutions in terms of special functions.

%%%%%%%%%%%%%%%%%%%%%%%%%%%%%%%%%%%%%%%%%
\section{Plane  limit}
\label{sec:PLimit}
%%%%%%%%%%%%%%%%%%%%%%%%%%%%%%%%%%%%%%%%%

In order to consider the plane limit $q\rightarrow0$ for $n$-point torus blocks in the necklace channel,  we expand the torus block with respect to the modular parameter as
\be
\label{asym}
\cF_{h,\td}(q, \bz) = q^{\td_1} \dps \sum^{\infty}_{p=0}\cF_p(\bz) q^p\,,
\ee
where the leading asymptotics  is always $q^{\td_1}$ due to the definition \eqref{gen_def}. In what follows we argue  that the expansion coefficient in the leading order $\cF_0(\bz)$ can be identified with a particular  conformal block  of the plane CFT$_2$. The exact relation is given by
\be
\label{med}
\cF_0(\bz)= \lim \limits_{\substack{%
\\
z_0 \to \infty\\
z_{n+1} \to 0}} z^{2\tilde h_1}_0 \tilde{\cF}(z_0, \bz, z_{n+1}) \,.
\ee
Here, $\tilde{\cF}(z_0, \bz, z_{n+1})$ is the $(n+2)$-point conformal block in the comb channel with external dimensions $(\td_1, h_1, h_2, \ldots, h_n, \td_1)$ and intermediate dimensions $(\tilde h_2, \tilde h_3, \ldots, \tilde h_n)$. In other words, the comb channel diagram is obtained from the necklace channel diagram by cutting the intermediate line of dimension $\td_1$ that produces two external lines of the same equal dimensions, see fig. \bref{Necklace}.

Let us demonstrate that $\tilde{\cF}(z_0, \bz, z_{n+1})$ satisfies the limiting Casimir equations \eqref{Casn1} which are exactly the Casimir equations in the comb channel of the plane CFT$_2$. Indeed,  substituting the expansion \eqref{asym} into the Casimir equation system \eqref{Casn1} and keeping the leading order in $q$ we find that the first equation in \eqref{Casn1} becomes  trivial, while the remaining $(n-1)$ equations in \eqref{Casn1} take the form
\be
\label{Clim}
\ba{c}
\dps \bigg[\tilde h_j(\tilde h_j-1)-\tilde h_1 (\tilde h_1 - 1) - \bigg(\sum_{k=1}^{j-1}\cL_{-1}^{(k)}\sum_{k=j}^{n}\cL_{1}^{(k)}\bigg) + \bigg(1 - 2 \tilde h_1 - \sum_{l=j}^n\cL_{0}^{(l)}\bigg)\sum_{k=j}^n\cL_{0}^{(k)}\bigg]\cF_0(\bz) = 0 \,,
\ea
\ee
where $j=2,...,n$.\footnote{The $1$-point torus block is determined only by the first equation in \eqref{Casn1} which is automatically satisfied at $q\rightarrow 0$. Thus, $n\geq 2$. The case $n=2$ is worked out in Appendix \bref{app:2pt_der}.} On the other hand, there are  $(n-1)$ Casimir equations for $(n+2)$-point conformal block in the comb channel that can be written as
\be
\label{C2n_pl}
\tilde{\mathbb{C}}_2[n+2,j] \tilde{\cF}(z_0, \bz, z_{n+1}) = -\tilde h_j(\tilde h_j-1)\tilde{\cF}(z_0, \bz, z_{n+1})\,,
\qquad
j = 2,...,n \,,
\ee
where the 2nd order differential operators on the left-hand side are directly read off from the $sl(2, \mathbb{R})$ Casimir operator \eqref{C2} as \cite{Dolan:2011dv,Alkalaev:2015fbw,Rosenhaus:2018zqn}:
\be
\label{Casn_flat}
\tilde{\mathbb{C}}_2[n+2,j]  = \mathbb{L}_{1}\mathbb{L}_{-1}+ \mathbb{L}_0-(\mathbb{L}_{0})^2\,,
\qquad
\quad
\mathbb{L}_{m} = \dps \sum_{k=j}^{n+1}\cL^{(k)}_{m}\,, \quad m = 0, \pm 1\,,
\ee
where the $sl(2, \mathbb{R})$ generators $\cL^{(k)}_{m}$ are given by \eqref{diff_L}. The above equations on the block function are supplemented by the Ward identities of the global $sl(2,\mathbb{R})$ (chiral) symmetry of the plane CFT$_2$:
\be
\label{ward_pl}
\sum_{k=0}^{n+1}\cL^{(k)}_{m} \tilde{\cF}(z_0, \bz, z_{n+1}) = 0\,,
\qquad
m = 0, \pm1\,.
\ee
Now, using the corollary of the Ward identity at $m=-1$ \eqref{ward_pl},
\be
\dps \sum_{k=j}^{n+1}\cL^{(k)}_{-1} \tilde{\cF}(z_0, \bz, z_{n+1}) = -\sum_{k=0}^{j-1}\cL^{(k)}_{-1} \tilde{\cF}(z_0, \bz, z_{n+1})\,,
\ee
we find that the plane Casimir equations \eqref{C2n_pl} can be cast into the form
\be
\label{cs}
\ba{c}
\dps \bigg[ \Big(- \sum_{k=1}^{j-1} \cL_{-1}^{(k)} \sum_{k=j}^{n}\cL_{1}^{(k)} + \sum_{k=j}^n\cL_{0}^{(k)} -\sum_{k=j}^n\cL_{0}^{(k)} \sum_{l=j}^n\cL_{0}^{(l)} - 2 \sum_{k=j}^n\cL_{0}^{(k)} \cL^{(n+1)}_{0} \Big) - \cL^{(0)}_{-1} \cL^{(n+1)}_{1} +
\vspace{2mm}
\\
\dps + \cL^{(n+1)}_{0} - \left(\cL^{(n+1)}_{0}\right)^2 - \sum_{k=1}^{j-1} \cL_{-1}^{(k)} \cL^{(n+1)}_{1} - \sum_{k=j}^{n}\cL_{1}^{(k)} \cL^{(0)}_{-1} + \tilde h_j(\tilde h_j-1) \bigg]\tilde{\cF}(z_0, \bz, z_{n+1}) = 0\,,
\ea
\ee
where we separated  groups of terms inside and outside the round brackets. In view of the relation \eqref{med} we recall that the comb channel block function  $\tilde{\cF}(z_0, \bz, z_{n+1})$ is regular at $z_{n+1} \rightarrow 0$ and grows as $z^{- 2 \tilde h_1}_{0}$ at $z_{0} \rightarrow \infty$. In particular, focusing on the terms outside the round brackets one  can shown the following asymptotic behaviour:
$$
\cL^{(n+1)}_{0} \tilde{\cF}(z_0, \bz, z_{n+1}) \Big|_{z_{n+1}\rightarrow 0} = \tilde h_1 \tilde{\cF}(z_0, \bz, 0) \,,
$$

\be
\label{ree}
\Big(\cL^{(n+1)}_{0} - \big(\cL^{(n+1)}_{0}\big)^2 \Big) \tilde{\cF}(z_0, \bz, z_{n+1}) \Big|_{z_{n+1}\rightarrow 0} = \tilde h_1(1 - \tilde h_1) \tilde{\cF}(z_0, \bz, 0) \,,
\ee

$$
\cL^{(n+1)}_{1} \tilde{\cF}(z_0, \bz, z_{n+1}) \Big|_{z_{n+1}\rightarrow 0} = 0\,,  \quad \dps \cL^{(0)}_{-1} \tilde{\cF}(z_0, \bz, z_{n+1}) \Big|_{z_{0} \rightarrow \infty} \sim z^{-2\tilde h_1-1}_0 \,.
$$
Taking the limit $z_0 \to \infty$ and $z_{n+1} \to 0$ in the plane Casimir equation \eqref{cs} and using the definition \eqref{med} along with the relations \eqref{ree} one finally arrives at the limiting torus  Casimir equations  \eqref{Clim}. Thus, this proves  the relation \eqref{med} between the leading $q\to 0$ asymptotics of the $n$-point  torus block $\cF_0(\bz)$ and the $(n+2)$-point plane block $\tilde{\cF}(z_0, \bz, z_{n+1})$.

Note that in the limit $q\to0$ the Casimir equations in the $\bm x$-parameterization \eqref{CasWithoutLeg} take  the form
\be
\label{Cas_x_pl}
\mathbb{C}^{^{q\to0}}_2[n,j]\, \cV_{0}(\bm x) = -\td_j(\td_j-1)\,\cV_{0}(\bm x)\,,
\qquad
j = 2,\ldots, n\,,
\ee
where the limiting  2nd order differential operators on the left-hand side are given by
\be
\label{C2n_pl2}
\mathbb{C}^{^{q\to0}}_2[n,j] = (1 - 2 \tilde h_1) N_j - N^2_j +\tilde h_1 (1 -\tilde h_1 ) - \sum_{k=1}^{j-1}\sum_{k=j}^{n} \cD_{i,k} \,,
\ee
and  $\cV_0(\bm x)$ is the leading asymptotics of the bare conformal block, which, therefore, defines the comb channel conformal block through the relations \eqref{leg} and  \eqref{med}. Equivalently, one can start with the limiting equations \eqref{Clim} and change variables ${\bf  z} \mapsto {\bm x}$.

Remarkably, the equations \eqref{Cas_x_pl} provide the explicit realization of the Casimir equations for multipoint conformal blocks of the plane  \cft in the comb channel. On the other hand, there is a nice parameterization in the plane CFT$_2$, in which $(n+2)$-point conformal blocks in the comb channel take the form \cite{Rosenhaus:2018zqn}
\be
\label{CombBlock}
\cV_0(\bm x) = \tilde{\mathbb{L}}(\bm x) \left(\prod_{i=1}^{n-1}\eta_i\right) F_K \left[
\begin{array}{l l}
\td_1 +\td_2 -h_1, \td_2+\td_3-h_2,\ldots, \td_n+\td_1-h_n\\
\qquad \qquad \qquad 2\td_2,\ldots, 2\td_n
\end{array}\bigg| \eta_1,\ldots, \eta_{n-1}
\right]\,,
\ee
where $F_K$ is the so-called comb function of the cross-ratios $\bm\eta = \{\eta_a({\bm x}), a = 1,...,n-1\}$  given by
\be
\eta_1 = x_2\frac{1-x_3}{1-x_2 x_3}\,,
\quad
\eta_i = x_{i+1}\frac{(1-x_i)(1-x_{i+2})}{(1-x_i x_{i+1})(1-x_{i+1} x_{i+2})}\,,
\quad
\eta_{n-1}=x_n\frac{1-x_{n-1}}{1-x_n x_{n-1}}\,,
\ee
while the leg factor reads
\be
\tilde{\mathbb{L}}(\bm x) = (1-x_2)^{\td_1-h_1}(1-x_n)^{\td_1-h_n}\prod_{j=2}^n \left(x_j^{-\td_1}\right) \prod_{i=2}^{n-1}\left(\frac{1-x_{i}x_{i+1}}{(1-x_{i})(1-x_{i+1})}\right)^{h_i}\,.
\ee

The comb function  satisfies simple recursive equations in $\bm \eta$ variables \cite{Rosenhaus:2018zqn}. Thus, we conclude that changing  variables $\bm x \mapsto \bm \eta$ the Casimir equations \eqref{Cas_x_pl} must take the form of the equations that determine the comb function.\footnote{It can be directly  checked for small $n$, although it is not quite clear how to prove this statement for any $n$.} However, using $\bm \eta$-parameterization instead of $\bm x$-parameterization  significantly complicates the torus Casimir equations \eqref{CasWithoutLeg}--\eqref{C2n}  beyond the $q\to 0$ limit. (In this respect, it would be crucial  to connect the eigenvalue torus Casimir  equations and their solutions  to Hamiltonians of integrable models and the theory of special functions \cite{Isachenkov:2016gim}, that, hopefully, will help to clarify the structure of torus conformal blocks.)

%%%%%%%%%%%%%%%%%%%%%%%%%%%%%%%%%%%%%%%%%
\section{Casimir equations in any channel}
\label{sec:any}
%%%%%%%%%%%%%%%%%%%%%%%%%%%%%%%%%%%%%%%%%

The Casimir equations for global torus blocks in arbitrary channels (see fig. \bref{f1} {\bf (b)}) can directly be built by using that a torus block is obtained by inserting projectors and acting with OPEs (see Section \bref{sec:block_diag}). To this end, we observe that the necklace sub-diagram satisfies the  same Casimir equations \eqref{Casn1}. Let us consider the matrix element \eqref{TorusFuncThroughPlaneElements} represented as $\Tr_{\td_1}\left[q^{L_0}\cO_1 \cO_2\ldots \cO_n\right]$,
where for simplicity we omitted $z$-dependence.
Then, one directly shows that the insertion of the Casimir operator between any pair of operators in the matrix element yields the same Casimir differential operator as in \eqref{Casn1_d}:
\be
\label{cas_main}
\Tr_{\td_1}\left[q^{L_0}\cO_1 \ldots \cO_{i-1}\mathbb{C}_2 \cO_{i} \ldots\cO_n\right]
 = \cC_2[n,i] \Tr_{\td_1}\left[q^{L_0}\cO_1 \ldots \cO_{i-1} \cO_{i} \ldots \cO_n\right],
\ee
where $i=1,...,n$. Indeed, the calculation is essentially the same as for the necklace channel blocks in Appendix \bref{app:2pt_der}. The only difference is that the necklace blocks contain projectors which, however, commute  with  $sl(2, \mathbb{R})$ generators $L_n$ (hence, with the Casimir operator $\mathbb{C}_2$). Thus, the resulting differential operators on the right-hand side remain the same no matter how many projectors are inserted in the matrix element on the left-hand side. The same holds true  for  OPE operators \eqref{OPE_oper} which also commute with $L_n$.

Note that the relations \eqref{cas_main} by no means are eigenvalue equations. This can be achieved by inserting a projector $\mathbb{P}_{\td}$ next to $\mathbb{C}_2$ in \eqref{cas_main} that produces an eigenvalue $-\td(\td-1)$ as
\be
\label{cas_main_P}
\ba{l}
\dps
\Tr_{\td_1}\left[q^{L_0}\cO_1 \ldots \cO_{i-1}\mathbb{P}_{\td}\mathbb{C}_2 \cO_{i} \ldots\cO_n\right]
 = \cC_2[n,i] \Tr_{\td_1}\left[q^{L_0}\cO_1 \ldots \cO_{i-1}\mathbb{P}_{\td} \cO_{i} \ldots \cO_n\right]
\vspace{2mm}
\\
\dps
\hspace{60mm} =
-\td(\td-1) \Tr_{\td_1}\left[q^{L_0}\cO_1 \ldots \cO_{i-1}\mathbb{P}_{\td}\cO_{i}\ldots\cO_n\right].
\ea
\ee
On the contrary, acting with OPE operators does not make eigenvalue equations from \eqref{cas_main}.  Since a particular combination of  $(k-1)$ projectors and $(n-k)$ OPEs singles out a torus block $\cF^{\,(k-1, \mathfrak{ch})}_{h,\td,c}(q,{\bf z})$ we conclude that the block function satisfies $(k-1)$  eigenvalue equations of the type \eqref{cas_main_P} while the rest $(n-k)$ eigenvalue equations are built as standard Casimir equations of plane CFT$_2$.

\begin{figure}%[H]
\centering
\includegraphics[width=0.45\linewidth]{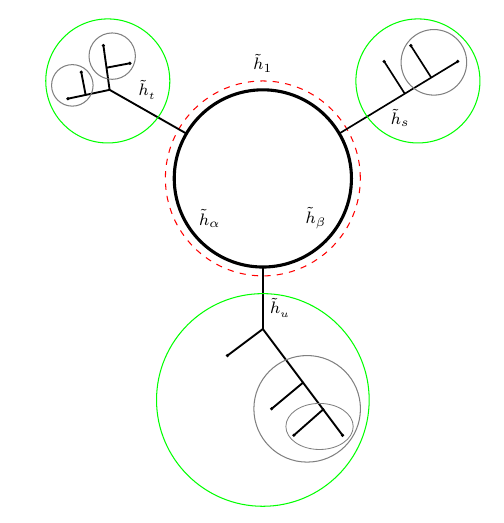}

\caption{The Casimir equations for $n$-point block $\cF^{\,(2, \mathfrak{ch})}_{h,\td,c}(q,{\bf z})$ with a three-leg necklace sub-diagram. The dashed red line symbolizes three Casimir equations associated to the necklace sub-diagram. The green contours indicate three Casimir equations for three internal dimensions of three segments composing the necklace. Each green contour contains smaller grey contours which define Casimir equations for other internal dimensions of planar sub-diagrams attached to the necklace.}
\label{mixed}
\end{figure}

As an example, consider inserting two projectors $\mathbb{P}_{\td_{\alpha}}$ and $\mathbb{P}_{\td_{\beta}}$ into the  $n$-point matrix element
\be
\label{1st_interP}
\Tr_{\td_1}\left[q^{L_0}\cO_1\ldots \cO_{i-1}\mathbb{P}_{\td_\alpha} \cO_{i}\ldots  \cO_{j-1}\mathbb{P}_{\td_\beta}\cO_j\ldots\cO_n\right]\,,
\qquad j>i\;,
\ee
that splits a string of $n$ operators into three subgroups, each of  which can be further fused by means of OPEs down to a single operator for each subgroup as
\be
\label{1st_interPOPE}
\Tr_{\td_1}\left[q^{L_0}\tilde\cO_s\,\mathbb{P}_{\td_\alpha} \tilde \cO_{t}\,\mathbb{P}_{\td_\beta}\tilde \cO_u\right],
\ee
where  $\tilde \cO$ have dimensions $\td_s, \td_t, \td_u$ and result from particular OPE orderings in each subgroup. In this way we obtain a torus block with a necklace sub-diagram having three legs connected to particular planar diagrams. Following the notation of Section \bref{sec:block_diag} this block is denoted $\cF^{\,(2, \mathfrak{ch})}_{h,\td,c}(q,{\bf z})$, where $\mathfrak{ch}$ describes OPE orderings in the three operator groups.

Then, using the relation \eqref{cas_main} we arrive at the following Casimir equations in the necklace sub-diagram
\be
\label{Casn_subneck}
\ba{c}
\cC_2[n, 1]\, \cF^{\,(2, \mathfrak{ch})}_{h,\td,c}(q,{\bf z}) = -\td_{1}(\td_1-1)\,\cF^{\,(2, \mathfrak{ch})}_{h,\td,c}(q,{\bf z})\,,
\vspace{2mm}
\\
\dps

\cC_2[n, i]\, \cF^{\,(2, \mathfrak{ch})}_{h,\td,c}(q,{\bf z}) = -\td_{\alpha}(\td_\alpha-1)\,\cF^{\,(2, \mathfrak{ch})}_{h,\td,c}(q,{\bf z})\,,
\vspace{2mm}
\\
\dps
\cC_2[n, j]\, \cF^{\,(2, \mathfrak{ch})}_{h,\td,c}(q,{\bf z}) = -\td_{\beta}(\td_\beta-1)\,\cF^{\,(2, \mathfrak{ch})}_{h,\td,c}(q,{\bf z})\,,
\ea
\ee
along with the plane Casimir equations describing OPE orderings $\mathfrak{ch}$ in each of three operator groups
\be
\label{Casn_legs}
\ba{l}
\dps
\tilde{\mathbb{C}}_2[1:i-1]\,\cF^{\,(2, \mathfrak{ch})}_{h,\td,c}(q,{\bf z}) = -\td_{s}(\td_s-1)\,\cF^{\,(2, \mathfrak{ch})}_{h,\td,c}(q,{\bf z})\,,
\vspace{2mm}
\\
\dps
\tilde{\mathbb{C}}_2[i:j-1]\,\cF^{\,(2, \mathfrak{ch})}_{h,\td,c}(q,{\bf z}) = -\td_{t}(\td_t-1)\,\cF^{\,(2, \mathfrak{ch})}_{h,\td,c}(q,{\bf z})\,,
\vspace{2mm}
\\
\dps
\tilde{\mathbb{C}}_2[j:n]\,\cF^{\,(2, \mathfrak{ch})}_{h,\td,c}(q,{\bf z}) = -\td_{u}(\td_u-1)\,\cF^{\,(2, \mathfrak{ch})}_{h,\td,c}(q,{\bf z})\,,
\ea
\ee
where similar to  \eqref{Casn_flat} we introduced the plane Casimir operators
\be
\label{Casn_flat1}
\tilde{\mathbb{C}}_2[a:b]  = \mathbb{L}_{1}\mathbb{L}_{-1}+ \mathbb{L}_0-(\mathbb{L}_{0})^2\,,
\qquad
\quad
\mathbb{L}_{m} = \dps \sum_{k=a}^{b}\cL^{(k)}_{m}\,, \quad m = 0, \pm 1\,.
\ee
Of course, this equation system \eqref{Casn_subneck}-\eqref{Casn_legs} is not complete because one has to add Casimir equations describing a particular channel in each of three groups of operators. This can be schematically  written as
\be
\tilde{\mathbb{C}}_2[A:B]\,\cF^{\,(2, \mathfrak{ch})}_{h,\td,c}(q,{\bf z}) = -\td_{C}(\td_C-1)\,\cF^{\,(2, \mathfrak{ch})}_{h,\td,c}(q,{\bf z})\,,
\ee
where $A,B$ are possible labels of primary operators in each subgroup of operators which on fig. \bref{mixed} correspond to the contours painted in grey, $\td_C$ stands for an intermediate dimension of the line crossed by a given contour.

The generalization to any number of projectors (legs of the necklace sub-diagram) is obvious. In the case of $n$-point global torus in the OPE channel (see fig. \bref{f1} {\bf (a)}) the above Casimir equations reproduce those ones given in \cite{Kraus:2017ezw}.

%%%%%%%%%%%%%%%%%%%%%%%%%%%%%%%%%%%%%%%%%
\section{Conclusion and outlooks}
\label{sec:conclusion}
%%%%%%%%%%%%%%%%%%%%%%%%%%%%%%%%%%%%%%%%%

In this paper we  elaborated on the conformal block decomposition  in the torus \cft and recognized  the  essential role of the necklace channel in building torus conformal blocks in general channels. We explicitly found  the Casimir equations  for the  multipoint  global torus block in the necklace channel \eqref{CasWithoutLeg}. The resulting PDE system is given by 2nd order differential operators both in the modular parameter and local coordinates. The conformal blocks diagonalize these operators with eigenvalues $-\td(\td-1)$ defined by dimensions of exchange channels. We have also presented the Casimir equations for global torus blocks in any channel.

In particular, the limiting  $q \rightarrow 0$ torus blocks with $n$ points in the necklace channel are recognized as plane blocks with $(n+2)$ points in the comb channel. Such an identification can be justified by simple counting of independent variables. The  $n$-point torus block depends on $(n-1)$ variables because of global $u(1)$ symmetry. At the same time, the $(n+2)$-point plane block  depends on the same $(n-1)$ variables because originally it depends on $(n+2)$ variables but 3 points can be fixed by the global $sl(2, \mathbb{R})$ symmetry. There is also a simple geometric argument underlying this relation: cutting a necklace  diagram yields a comb diagram. As a by-product, we obtained the plane Casimir equations as explicit 2nd order differential operators in given coordinates and conformal blocks are their eigenfunctions \eqref{Cas_x_pl}.

Our study of global torus conformal blocks was partly motivated by Wilson network realization of conformal blocks in the context of \adscft correspondence \cite{Bhatta:2016hpz,Fitzpatrick:2016mtp,kraus_new,Kraus:2017ezw,Hikida:2018dxe,Hikida:2018eih,Alkalaev:2020yvq,Castro:2020smu,Belavin:2022bib}. On the other hand, semiclassical heavy-light torus blocks dual to geodesic networks in the thermal \ads can be obtained from the large-$h$ (``heavy-light'') expansion of global torus blocks \cite{Fitzpatrick:2015zha,Alkalaev:2015fbw,Alkalaev:2016fok,Alkalaev:2016ptm}. The Casimir approach could be useful here, see e.g. \cite{Kraus:2017ezw}.

 Also, from the \adscft perspective it would be useful to study  $\cN=1$ supersymmetric extensions of the Casimir approach for $n$-point superblocks similar to the case of 1-point torus superblocks \cite{Alkalaev:2018qaz}.

%\vspace{2mm}

\paragraph{Acknowledgements.} Our work was supported by the Foundation for the Advancement of Theoretical Physics and Mathematics “BASIS”.

%%%%%%%%%%%%%%%%%%%%%%%%%%%%%%%%%%%%%%%%%
\appendix

\section{Derivation of the torus  Casimir equations for 2-point blocks}
\label{app:2pt_der}
%%%%%%%%%%%%%%%%%%%%%%%%%%%%%%%%%%%%%%%%%

In what follows we explicitly derive the Casimir equations in the  case $n=2$ and analyze their limits $n=2 \to  n=1$ at any $q$  and $q\to0$ at $n=2$.

\paragraph{Derivation of 2-point Casimir equations.} In the 2-point case, the  Casimir irreducibility conditions \eqref{Cas_eq_n} are given by two relations
\begin{equation}
\label{listOfEqs}
\ba{c}
\dps
 1)\;\; \Tr_{\td_1} [\C2 q^{L_0}\cO_1 \mathbb{P}_{\td_2 } \cO_2] = -\td_1 (\td_1 -1) \cF_{h,\td}\,,
\vspace{2mm}
\\
\dps
2)\;\; \Tr_{\td_1} [q^{L_0}\cO_1 \mathbb{P}_{\td_2 }\C2 \,\cO_2] = -\td_2 (\td_2 -1) \cF_{h,\td}\,.
\ea
\end{equation}
In order to represent   \eqref{listOfEqs} as the eigenvalue equations for the 2-point block function $\cF_{h,\td}$ in the necklace channel \eqref{neck_block_n}  one substitutes the Casimir operator $\operatorname{C_2} = L_1 L_{-1} - L_0 - L_0^2$ \eqref{C2} into the both relations. Each of the three terms in $\operatorname{C_2}$ is to be  treated  separately.

Let us begin with the first relation in \eqref{listOfEqs} and consider the term  $L_1L_{-1}$. The trick is to commute $L_{-1}$ to the right to obtain the following relation
\begin{equation}
\label{casimir12}
\begin{split}
\Tr_{\td_1} [L_1 L_{-1} q^{L_0}\cO_1 \mathbb{P}_{\td_2 } \cO_2] &= \frac{1}{q}\Tr_{\td_1} [L_1 q^{L_0} L_{-1} \cO_1 \mathbb{P}_{\td_2 } \cO_2] =\\
&= \frac{1}{q} \cL_{-1}^{(1)}\Tr_{\td_1} [L_1 q^{L_0} \cO_1 \mathbb{P}_{\td_2 } \cO_2] + \frac{1}{q}\Tr_{\td_1} [L_1 q^{L_0} \cO_1 \mathbb{P}_{\td_2 } L_{-1} \cO_2] \,,
\end{split}
\end{equation}
where we used  $L_n q^{L_0} = q^{L_0+n}L_n$, $L_n \mathbb{P} = \mathbb{P} L_n$, and  $[L_n,\cO_i(z_i)] = \cL_{n}^{(i)}\cO_i(z_i)$ \eqref{diff_L}. Commuting $L_{-1}$ with   $\cO_2$ in the last trace in \eqref{casimir12}, using the cyclic property of the trace and then commuting operators $L_{\pm1}$, we obtain
\begin{equation}
\label{interexpL1m1}
\ba{l}
\dps
\Tr_{\td_1} [L_1 L_{-1} q^{L_0}\cO_1 \mathbb{P}_{\td_2 } \cO_2] = \frac{1}{q} (\cL_{-1}^{(1)}+\cL_{-1}^{(2)})\Tr_{\td_1} [L_1 q^{L_0} \cO_1 \mathbb{P}_{\td_2 } \cO_2]
\vspace{2mm}
\\
\dps
\hspace{43mm}-\frac{2}{q}\Tr_{\td_1}[L_0q^{L_0}\cO_1 \mathbb{P}_{\td_2 } \cO_2]
+ \frac{1}{q}\Tr_{\td_1} [L_1 L_{-1} q^{L_0}\cO_1 \mathbb{P}_{\td_2 } \cO_2]\,.
\ea
\end{equation}
The last step is to calculate the trace $\Tr_{\td_1} [L_1 q^{L_0} \cO_1 \mathbb{P}_{\td_2 } \cO_2]$ in the first line of \eqref{interexpL1m1}. Using the same procedure one finds
\begin{equation}
\begin{split}
\Tr_{\td_1}[L_1 q^{L_0} \cO_1 \mathbb{P}_{\td_2 } \cO_2] &= q \left(\cL_{1}^{(1)} + \cL_{1}^{(2)}\right)\cF_{h,\td} + q \Tr_{\td_1}[q^{L_0} \cO_1 \mathbb{P}_{\td_2 } \cO_2 L_1] \,.
\end{split}
\end{equation}
By the cyclic property of the trace the last term is similar to the left-hand side term. Substituting it back into \eqref{interexpL1m1} one obtains
\begin{equation}
\begin{split}
\Tr_{\td_1} [L_1 L_{-1} q^{L_0}\cO_1 \mathbb{P}_{\td_2 } \cO_2] &= \frac{2}{1-q}q\partial_q \cF_{h,\td} - \frac{q}{(1-q)^2}\left(\cL_{-1}^{(1)}+\cL_{-1}^{(2)}\right)\left(\cL_{1}^{(1)} + \cL_{1}^{(2)}\right)\cF_{h,\td}\,.
\end{split}
\end{equation}
Now, $L_0$ and $L_0^2$ terms in the Casimir operator are calculated by using the following  obvious relations
\be
\Tr_{\td_1} [L_0 q^{L_0}\cO_1 \mathbb{P}_{\td_2 } \cO_2] = q \partial_q \cF_{h,\td}\,,
\qquad
\Tr_{\td_1} [L^2_0 q^{L_0}\cO_1 \mathbb{P}_{\td_2 } \cO_2] = (q \partial_q)^2 \cF_{h,\td}\,.
\ee
Finally, the first Casimir equation is cast into the form
\begin{equation}
\label{1Casimir}
\begin{split}
\left[-q^2\partial_q^2 + \frac{2q}{1-q}q\partial_q - \frac{q}{(1-q)^2}\left(\cL_{-1}^{(1)}+\cL_{-1}^{(2)}\right)\left(\cL_{1}^{(1)} + \cL_{1}^{(2)}\right)\right] \cF_{h,\td} = -\td_1 (\td_1 -1)\cF_{h,\td}\,,
\end{split}
\end{equation}
which is now the eigenvalue equation for  $\cF_{h,\td}$. This is the first equation in \eqref{Casn1} at $n=2$.

The same procedure applies to  the second relation in \eqref{listOfEqs}. Isolating the term $L_{1}L_{-1}$ in the Casimir operator \eqref{C2} one obtains
\begin{equation}\label{interExp2}
\begin{split}
\Tr_{\td_1} [q^{L_0}\cO_1 \mathbb{P}_{\td_2 }L_1 L_{-1}\cO_2] &=\frac{1}{q}(\cL_{-1}^{(1)}+q\cL_{-1}^{(2)})\Tr_{\td_1}[q^{L_0}\cO_1 \mathbb{P}_{\td_2 }L_1\cO_2]-\\
&-\frac{2}{q} \Tr_{\td_1}[q^{L_0}\cO_1 \mathbb{P}_{\td_2 }L_0\cO_2]+\frac{1}{q}\Tr_{\td_1} [q^{L_0}\cO_1 \mathbb{P}_{\td_2 }L_1 L_{-1}\cO_2]\,,
\end{split}
\end{equation}
where  first and  second traces on the right-hand side can be cast into the form
\begin{equation}
\label{L0for2Casimir}
\ba{l}
\dps
 \Tr_{\td_1}[q^{L_0}\cO_1 \mathbb{P}_{\td_2 }L_1\cO_2] = \frac{1}{1-q}\left(\cL_1^{(2)}+q\cL_1^{(1)}\right)\cF_{h,\td}\,,
\vspace{2mm}
\\
\dps
\Tr_{\td_1}[q^{L_0}\cO_1 \mathbb{P}_{\td_2 }L_0\cO_2] = \cL_0^{(2)}\cF_{h,\td} + q\partial_q \cF_{h,\td}\,.
\ea
\end{equation}
Substituting these expressions into \eqref{interExp2} yields
\begin{equation}\label{interExpL1Lm1Second}
\begin{split}
\Tr_{\td_1} [q^{L_0}\cO_1 \mathbb{P}_{\td_2 }L_1 L_{-1}\cO_2] &= \frac{2}{(1-q)}q\partial_q \cF_{h,\td} + \frac{2}{(1-q)}\cL_0^{(2)}\cF_{h,\td} -\\
&-\frac{1}{(1-q)^2}\left(\cL_{-1}^{(1)}+q\cL_{-1}^{(2)}\right)\left(\cL_1^{(2)}+q\cL_1^{(1)}\right)\cF_{h,\td}\,.
\end{split}
\end{equation}
Using \eqref{L0for2Casimir} one derives the last term in Casimir operator \eqref{C2} as
\begin{equation}\label{L02for2Casimir}
\begin{split}
\Tr_{\td_1}[q^{L_0}\cO_1 \mathbb{P}_{\td_2 }L_0^2\cO_2] = \left(\cL_0^{(2)}\right)^2\cF_{h,\td} + 2\cL_0^{(2)}q\partial_q \cF_{h,\td} + q\partial_q(q\partial_q \cF_{h,\td})\,.
\end{split}
\end{equation}
Substituting \eqref{interExpL1Lm1Second}, \eqref{L0for2Casimir}, \eqref{L02for2Casimir} into \eqref{listOfEqs} one finally obtains
\begin{equation}
\label{2Casimir}
\ba{l}
\dps
\left[- q^2\partial_q^2+\frac{2q}{1-q}q\partial_q - \frac{1}{(1-q)^2}\left(\cL_{-1}^{(1)}+q\cL_{-1}^{(2)}\right)\left(\cL_1^{(2)}+q\cL_1^{(1)}\right)
\right.
\vspace{2mm}
\\
\dps
\hspace{40mm}\left. +\frac{1+q}{1-q}\cL_0^{(2)} -\left(\cL_0^{(2)}\right)^2 - 2\cL_0^{(2)}q\partial_q  \right] \cF_{h,\td}
= -\td_2 (\td_2 -1)\cF_{h,\td}\,,
\ea
\end{equation}
which  is the second equation in \eqref{Casn1} at $n=2$. It is straightforward to check that the 2-point torus block function \eqref{nglobs} does solve the Casimir equation system \eqref{1Casimir} and  \eqref{2Casimir}. The same procedure in the $n$-point case yields the Casimir  system \eqref{Casn1}.

\paragraph{Reduction to the 1-point Casimir equation.}  Let us show that the 2-point  Casimir equations \eqref{1Casimir} and \eqref{2Casimir} are reduced to the 1-point Casimir equation \cite{Kraus:2017ezw}. To this end, we choose the second primary operator to be the identity operator, $\cO_2(z_2) = \mathbb{I}$, i.e. $h_2=0$. It follows that the intermediate dimensions must be equated, $\td_1 =\td_2$.

The 2-point conformal block $\cF_{h, \td} \equiv \cF_{h_1h_2, \td_1 \td_2}$ satisfies the Ward identity \eqref{nWard}:  $(\cL_0^{(1)}+\cL_0^{(2)})\cF_{h_1h_2, \td_1 \td_2} = 0$,
which by using \eqref{diff_L} takes the form
\begin{equation}
\label{WardExpl}
\left(h_1+h_2 + z_1\partial_1 + z_2 \partial_2\right)\cF_{h_1h_2, \td_1 \td_2} = 0\,.
\end{equation}
On the other hand, the $1$-point block $\cF_{h_1, \td_1} \equiv \cF_{h_1h_2, \td_1 \td_2}\big{|}_{h_2=0,\td_2=\td_1}$ satisfies the   Ward identity $\cL_0^{(1)}\cF_{h_1,\td_1} = 0$. Then, together with the condition $h_2=0$, equation \eqref{WardExpl} gives
\begin{equation}\label{1pointCond}
\partial_2 \cF_{h_1,\td_1} = 0\,.
\end{equation}
Thus, the first Casimir equation \eqref{1Casimir} takes the form \cite{Kraus:2017ezw}
\begin{equation}
\label{1pointCas}
\left[-q^2\partial_q^2 +\frac{2q}{1-q}q\partial_q+\frac{q}{(1-q)^2}h_1(h_1-1)\right] \cF_{h_1,\td_1}=-\td_1 (\td_1 -1)\cF_{h_1,\td_1}\,.
\end{equation}
Similarly, using  \eqref{1pointCond} and $\td_1 =\td_2 $ one shows that the second Casimir equation \eqref{2Casimir} also reduces to \eqref{1pointCas}.

\paragraph{Reduction from the $2$-point torus block to the $4$-point plane block.} Let us illustrate the relation \eqref{med} for the $2$-point torus block with external dimensions $(h_1, h_2)$ and intermediate dimensions $(\td_1, \td_2)$ in the plane limit. Setting $n=j=2$ in \eqref{Clim} we find
\be
\label{2ptl}
\bigg[\tilde h_1 (1 -\tilde h_1 ) -\cL_{-1}^{(1)}\cL_{1}^{(2)} +(1 - 2 \tilde h_1)\cL_{0}^{(2)}-\left(\cL_{0}^{(2)}\right)^2 + \tilde h_2 (\tilde h_2 -1)\bigg] \cF_{0}(z_1, z_2) = 0\,.
\ee
Using \eqref{diff_L} one directly shows that \eqref{2ptl} is reduced to the hypergeometric equation and its solution has the following form
\be
\label{sol_4}
\cF_{0}(z_1, z_2) = z_1^{-h_1} z_2^{-h_2}\; x^{\td_2 -\td_1} \;{}_2F_1 \left(\td_2 -\td_1 +h_1,\td_2 -\td_1 +h_2, 2\td_2\, \big| \,x\right),
\ee
where $x = z_2/z_1$.  On the other hand, the 4-point plane block  is known to be given by the hypergeometric function of external dimensions $(h_0, h_1, h_2, h_3)$ and intermediate dimension $\td$:
\be
\label{4pt_pl}
\ba{l}
\dps
\tilde \cF(z_0,z_1, z_2, z_3| h, \td) = z_{01}^{-h_0-h_1}z_{23}^{-h_2 - h_3} z_{02}^{-h_{0}+h_{1}}z_{12}^{h_{0}-h_{1}-h_{2}+h_{3}}z_{13}^{h_{2}-h_{3}}
\vspace{2mm}
\\
\dps
\hspace{30mm}\times\,  \eta^{\tilde h} (1-\eta)^{h_2-h_3-h_0+h_1} \;{}_2F_1 \left( \tilde h - h_{0}+h_1, \tilde h +h_{2}-h_{3}, 2 \tilde h\,\big|\, \eta\right),
\ea
\ee
where $z_{ij} \equiv z_i - z_j$ and the cross-ratio $\eta = (z_{01} z_{23})/(z_{02} z_{13})$.  Then, \eqref{sol_4} is indeed related to \eqref{4pt_pl} by \eqref{med}  provided that the external/intermediate  dimensions in \eqref{4pt_pl} are chosen as  $(\td_1, h_1, h_2, \td_1)$ and  $\td_2$.

\section{Explicit expressions for lower-point global torus block}
\label{app:block}

Here we collect a few explicit block functions.

\begin{itemize}

\item The $1$-point block reads \cite{Hadasz:2009db}
\be
\label{expl_1}
\cF_{h,\td}(q,z) =  z^{-h} \sum^{\infty}_{n=0} \frac{\tau_{n,n}(\tilde h, h, \tilde h)}{n! (2\tilde h)_n} q^{\tilde h +n},
\ee
where $\tau$-coefficients are given by \eqref{TauElement}.

\item The $2$-point block in the necklace channel is (here and below $x_j = z_j/z_{j-1}$) \cite{Alkalaev:2017bzx}
\be
\label{expl_2}
\cF_{h_1 h_2,\td_1,\td_2}(q, z_1, z_2) = z^{-h_1}_1 z^{-h_2}_2 \sum^{\infty}_{n,m=0} \frac{\tau_{n,m}(\tilde h_1, h_1, \tilde h_2) \tau_{m,n}(\tilde h_2, h_2, \tilde h_1)}{n!m! (2\tilde h_1)_n (2 \tilde h_2)_m} q^{\tilde h_1  +n} x_2^{\tilde h_2 - \tilde h_1 + m- n}.
\ee
\item Using   \eqref{expl_1} one can find the 2-point and 3-point OPE blocks shown on fig. \bref{fig_23pt} {\bf (c)}  and {\bf (f)} given by \eqref{fex} and \eqref{nex}, respectively \cite{Kraus:2017ezw,Alkalaev:2017bzx}:
\be
\cF^{\,(0,\mathfrak{ch}_1)}_{h_1 h_2, \td_1 \td_2} (q, z_1, z_2) =  z^{-h_1}_1 z^{-h_2}_2  \sum^{\infty}_{n,m=0} \frac{\tau_{n,n}(\tilde h_1, \tilde h_2, \tilde h_1) \sigma_{m}}{n!m! (2\tilde h_1)_n (2 \tilde h_2)_m} q^{\tilde h_1  +n} (1-x_2)^{\tilde h_2 - h_1 -h_2 +m} x^{h_2 - \tilde h_2 -m}_2,
\ee
where $\sigma_m = (-1)^m (\tilde h_2)_m (\tilde h_2 +h_1 - h_2)_m$, and
\be
\ba{c}
\cF^{\,(0,\mathfrak{ch}_2)}_{h_1 h_2 h_3,\td_1 \td_2 \td_3}(q,z_1, z_2, z_3)  = z^{-h_1}_1 z^{-h_2}_2 z^{-h_3}_3 \dps  \sum^{\infty}_{n,m,k=0}   \frac{\tau_{n,n}(\tilde h_1, \tilde h_2, \tilde h_1) \sigma_{m,k}}{n!m!k! (2\tilde h_1)_n (2 \tilde h_2)_m (2 \tilde h_3)_k}  q^{\tilde h_1  +n}
\\
 \times (1-x_2)^{\tilde h_3 - h_1 - h_2 +k} (1-x_3)^{\tilde h_2 - h_3 - \tilde h_3 +m -k} x^{h_2 - \tilde h_3 -k}_2 x^{-\tilde h_2 -m}_3,
\ea
\ee
where $\sigma_{m,k} = (-1)^{m+k} (\tilde h_2)_m (\tilde h_3 +h_1 - h_2)_k (\tilde h_2 + \tilde h_3 - h_3+m)_k (\tilde h_2 + \tilde h_3 - h_3)_m$.
\item The mixed channel 3-point block (see fig. \bref{fig_23pt} {\bf (e)})  is determined by \eqref{expl_2} and \eqref{nexx}
\be
\ba{c}
\cF^{\,(1, \mathfrak{ch}_1)}_{h_1 h_2 h_3,\td_1 \td_2 \td_3}(q,z_1, z_2, z_3) = z^{-h_1}_{1} z^{-h_2}_2 z^{-h_3}_3 \dps \sum^{\infty}_{m,n,k =0} \frac {\tau_{m,k} (\tilde h_1, \tilde h_3, \tilde h_2) \tau_{k,m} (\tilde h_2, h_3, \tilde h_1) \sigma_{m,n,k}}{m!n!k! (2\tilde h_1)_m (2\tilde h_2)_k (2\tilde h_3)_n}
\\
\dps \times q^{\tilde h_1 + m} (1- x_2)^{\tilde h_3 -h_1 - h_2 +n} x^{h_2 - \tilde h_3}_2 x_3^{\tilde h_2 - \tilde h_1 + k - m}\;,
\ea
\ee
where $\sigma_{m,n,k}= (-1)^n(\tilde h_1 - \tilde h_3 - \tilde h_2 + m-k-n+1)_n (\tilde h_3 + h_2 -h_1)_n$. It is a straightforward to check that this function satisfies the Casimir equations in the mixed channel \eqref{Casn_subneck}-\eqref{Casn_legs}.

\end{itemize}

%\bibliographystyle{JHEP}
%\bibliography{refs}

\providecommand{\href}[2]{#2}\begingroup\raggedright\endgroup

\end{document}